\documentclass[11pt]{article}
\usepackage{amsmath,color}
\usepackage{amssymb}
\usepackage{amsfonts}
\usepackage{graphicx}
\usepackage{color}
\definecolor{darkgreen}{rgb}{0,0.35,0}

\numberwithin{equation}{section}

\begin{document}

\title{\textbf{Black hole solutions in Chern-Simons AdS supergravity}}
\author{Gaston Giribet\medskip\\
\textit{Departamento de F\'{\i}sica, Universidad de Buenos Aires FCEN-UBA, }\\
\textit{IFIBA-CONICET, Ciudad Universitaria, }\\
\textit{Pabell\'{o}n I, 1428, Buenos Aires, Argentina}\\
{\small E-mail: gaston-at-df.uba.ar}\medskip
\and Nelson Merino and Olivera Miskovic\medskip\\
\textit{Instituto de F\'{\i}sica, Pontificia Universidad Cat\'{o}lica de Valpara\'{\i}so, }\\
\textit{Casilla 4059, Valpara\'{\i}so, Chile}\\
{\small E-mail: nelson.merino-at-ucv.cl, olivera.miskovic-at-ucv.cl}
\and Jorge Zanelli\medskip\\
\textit{Centro de Estudios Cient\'{\i}ficos (CECs), Av.Arturo Prat 514, Valdivia, Chile}\\
\textit{Universidad Andr\'es Bello, Av. Rep\'ublica 440, Santiago, Chile}\\
{\small E-mail: z-at-cecs.cl}\medskip}

\maketitle

\begin{abstract}
We study charged AdS black hole solutions in five-dimensional Chern-Simons supergravity. The minimal supergroup containing such AdS$_5 \times U(1)$ configurations is the superunitary group $SU(2,2|{\mathcal N})$. For this model, we find analytic black hole solutions that asymptote to locally AdS$_5$ spacetime at the boundary. A solution can carry $U(1)$ charge provided the spacetime torsion is non-vanishing. Thus, we analyze the most general configuration consistent with the local AdS$_5$ isometries in Riemann-Cartan space. The coupling of torsion in the action resembles that of the universal axion of string theory, and it is ultimately due to this field that the theory acquires propagating degrees of freedom. Through a careful analysis of the canonical structure the local degrees of freedom of the theory are identified in the static symmetric sector of phase space.
\end{abstract}

%%%%%%%%%%%%%%%%%%%
\section{Introduction}             %  1  %
%%%%%%%%%%%%%%%%%%%

In the framework of the AdS/CFT correspondence \cite{Maldacena, Witten, AdS/CFT}, asymptotically locally AdS$_5$ black holes represent the gravitational configurations dual to conformal field theories at finite temperature. In this setup, charged AdS black holes are specially relevant to address a vast variety of problems essential to describe phenomena like thermalization in the presence of chemical potentials and superconducting phases, among others.

In AdS/CFT, a crucial role is played by local symmetries. Local symmetries in the bulk correspond to global symmetries at the boundary, and thus symmetry breaking in the bulk induces quantum anomalies in the dual CFT. In this context, if one is led by the gauge invariance principle to build a sensible gravitational theory, it is quite natural to investigate the case of Chern-Simons (CS) supergravity theories. Among the attractive features of CS supergravity in AdS, we find that in such a setup the graviton, the bosonic matter, and the fermions enter on equal footing in the action, all of them being different components of the same connection for a supergroup that contains the AdS isometry group, the internal gauge symmetry group, and the supersymmetry transformations.

An additional motivation to consider pure CS gravity theories comes from the fact that they belong to the class of Lovelock theories, which provide the natural generalization of General Relativity for higher dimensions. At the point in parameter space of the five-dimensional Lovelock theory that corresponds to CS gravity, the local symmetry of the theory is enhanced from $SO(4,1)$ (Lorentz group)  to $SO(4,2)$ (the AdS$_5$). CS supergravities with local symmetries that contain AdS groups are genuine gauge theories of gravity and are known in all odd dimensions Refs.\cite{Chamseddine,Banados-Troncoso-Zanelli,Troncoso-Zanelli}.

Here we focus on the five-dimensional asymptotically locally AdS spacetimes whose three-dimensional constant-radius section is maximally symmetric. The interest in this particular example is that in such case the boundary corresponds to four-dimensional flat space, which is the physically interesting case for holographic applications. Besides, five is the smallest dimension in which a CS supergravity model contains a propagating graviton \cite{Banados-Garay-Henneaux}, in contrast to the 3D case where CS gravity is a topological theory with no local degrees of freedom \cite{Achucarro-Townsend}.

The minimal content of physical fields necessary to have a charged black hole solution in AdS$_5$ CS supergravity, contains the $SO(4,2)$ gauge field associated to the graviton \cite{SO(4-2)}, and an Abelian gauge $U(1)$ field that introduces electromagnetic interaction. The smallest supersymmetric extension of AdS$_5\times U(1)$ is the supergroup $SU(2,2|{\mathcal N})$, which in addition contains non-Abelian $SU({\mathcal N})$ interaction, and fermions. The gauge connection 1-form $\mathbf{A}$ for this supergroup has associated field-strength 2-form $\mathbf{F}=d\mathbf{A+A\wedge A}$. A gauge-invariant quantity constructed from $\mathbf{F}$ is the trace $\frac{i}{3}\text{Tr}\,(\mathbf{F\wedge F\wedge F})=dL_{\text{CS}}$ which is an exact six-form that can be locally written as the exterior derivative of a Chern-Simons five-form. This CS from defines a five-dimensional Lagrangian density, $L_{\text{CS}}(\mathbf{A})$, which is gauge-invariant up to a boundary term. This formalism naturally describes a Riemann-Cartan spacetime, where curvature and torsion enter in the AdS components of the gauge supergroup field-strength.

In order to study charged black holes, torsion, fermions and non-Abelian gauge fields can be switched off, keeping only the metric (or vielbein) and the Abelian connection as fundamental fields. The fermions can be added later to study the stability of the solution through its preserved supersymmetries. This technique has been applied to CS supergravity for instance in Ref. \cite{Miskovic-Troncoso-Zanelli}, where a global AdS solution containing Abelian matter with non-trivial winding was shown to be stable due to some supersymmetries that remain unbroken. In this work, we are interested in a symmetric ansatz, in which the static black hole metric possesses maximal number of isometries and is charged only under the Abelian field. As shown below, there are no charged black hole solutions because gravity decouples from the $U(1)$ field, unless the spacetime has torsion. Indeed, in order to couple electromagnetism and gravity torsion is needed, and the way torsion enters in the CS action resembles the so-called universal axion of string theory (see, for example, Ref. \cite{Green-Schwarz-Witten}).

The paper is organized as follows: In Section 2, the CS gravity of interest is reviewed and its field equations are presented. Section 3 discusses the static black hole solutions with locally flat horizon, and our results are compared with other found in the literature in Section 4. In Section 5 it is shown that the most general solution exhibits properties that are a consequence of additional
local symmetries in the theory. A proof of this claim is given in Section 6 using the Hamiltonian analysis. Finally, Section 7 contains the conclusions.

%%%%%%%%%%%%%%%%%%%%%%%%%%%%%%%%%%%%%%%%
\section{Chern-Simons AdS supergravity in AdS$\times U(1)$ sector}  %  2  %
%%%%%%%%%%%%%%%%%%%%%%%%%%%%%%%%%%%%%%%%

Electrically charged AdS black holes in five-dimensional Chern-Simons (CS) supergavity \cite{Chamseddine,Troncoso-Zanelli} can be obtained from the AdS$\times U(1)$ sector of this theory, that is, when the fermions and non-Abelian bosons are switched off. The full CS supergravity action is given in Appendix \ref{CS SUGRA}. Then, the Lie algebra-valued gauge connection 1-form is
\begin{equation}
\mathbf{A}=\frac{1}{2}\,\omega^{ab}\mathbf{J}_{ab}+\frac{1}{\ell}\,e^{a}\mathbf{J}_{a}+A\,\mathbf{T}_1\,,\label{A}
\end{equation}
where $ \mathbf{J}_{ab},\mathbf{P}_a,\mathbf{T}_1$ are the anti-Hermitean generators whose algebra is $\mathfrak{so}(4,2) \oplus \mathfrak{u}(1)$. Here $\mathbf{J}_{ab}=-\mathbf{J}_{ba}$\ $(a=0,\ldots,4)$ and $\mathbf{J}_a$, generate Lorenz rotations and AdS boosts, respectively, and $\mathbf{T}_1$ is the Abelian generator. When the five-dimensional bulk manifold $\mathcal{M}$ is parametrized by the local coordinates $x^\mu$, the fundamental fields in (\ref{A}) are the vielbein $e^a=e_\mu^a(x)\,dx^\mu$, spin connection $\omega^{ab}=\omega_{\mu}^{ab}(x)\,dx^{\mu}$ and electromagnetic gauge field $A=A_{\mu}(x)\,dx^{\mu}$. The associated field strength,
\begin{equation}
\mathbf{F}=\frac{1}{2}\,F^{ab}\mathbf{J}_{ab}
+\frac{1}{\ell}\,T^{a}\mathbf{J}_{a}+F\,\mathbf{T}_1\,,
\end{equation}
is related to the Lorentz curvature 2-form $R^{ab}=d\omega^{ab}+\omega_{\ c}^a\wedge\omega^{cb}$ through
\begin{equation}
F^{ab}=R^{ab}+\frac{1}{\ell^{2}}\,e^{a}\wedge e^b\,,
\end{equation}
and the spacetime torsion 2-form is $T^a=De^a=de^a+\omega^{a}_b\wedge e^b$, with gauge group covariant derivative $D=d+\left[\omega,\quad\right]$. The Abelian field strength is $F=dA$.

The CS Lagrangian for AdS gravity in five dimensions can be implicitly defined in a gauge-invariant way as
\begin{equation}
dL_\text{CS}=\frac{i}{3}\left\langle \mathbf{F}^3 \right\rangle_{g}=\frac{1}{3}\,g_{MNK}\,F^M\wedge F^N\wedge F^K\, .
\end{equation}
Here $\left\langle \mathbf{\ldots}\right\rangle _g$ is defined by the symmetric invariant tensor $g_{MNK}=i\left\langle \mathbf{T}_M \mathbf{T}_N \mathbf{T}_K \right\rangle_g$, where the generators are collectively denoted $\mathbf{T}_M= \left\{\mathbf{J}_{ab},\mathbf{P}_a,\mathbf{T}_1\right\}$. {The most general form of this invariant tensor has all components non-vanishing, except of $g_{11M\neq 1}=0$. The Cartan metric $\left\langle \mathbf{T}_M\mathbf{T}_K \right\rangle _g$ can always be chosen flat and the invariant tensor of AdS group that exists in any odd dimension is given by the completely antisymmetric tensor. Therefore, non-vanishing components of the invariant tensor can be written as }
\begin{align}
g_{a[bc] [de]} &  =k\,\epsilon_{abcde}\,,\nonumber\\
g_{1 [ab] [cd] } &  ={\alpha}\,\left(  \eta_{ac}\eta_{bd}-\eta_{ad}\eta_{bc}\right)  \,,\nonumber\\
g_{1ab} &  =-{\alpha} \,\eta_{ab}\,,\nonumber\\
g_{111} &  = {\beta} \,,\label{g}
\end{align}
where {$k,\alpha$\ and $\beta$} are real constants, and $[ab], [cd],\ldots$ are pairs of antisymmetrized indices. In our notation, the signature is $\eta_{ab}=\,$diag$(-,+,+,+,+) $.

Dropping the wedge product for the sake of simplicity, the CS action can be written as
\begin{align}
I_\text{CS}[\mathbf{A}]   & =\int_{M}L_\text{CS}(\mathbf{A})=\frac{i}{3}\int\limits_{M}\left\langle \mathbf{AF}^{2}
-\frac{1}{2}\,\mathbf{FA}^{3}+\frac{1}{10}\,\mathbf{A}^{5}\right\rangle \nonumber\\
&  =\int\limits_{M}\left[  L_{\text{AdS}}(e,\omega)
+L_{U(1)}(A)+L_{\text{int}}(e,\omega,A)\right]  \,,\label{CS}
\end{align}
where the pure AdS and $U(1)$ CS Lagrangians read
\begin{align}
L_{\text{AdS}}(e,\omega) &  =\frac{k}{4\ell}\,\epsilon_{abcde}\left(R^{ab}R^{cd}+\frac{2}{3\ell^2}\,R^{ab}e^c e^d + \frac{1}{5\ell^4}\,e^a e^b e^c e^d\right) e^e \,,\nonumber\\
L_{U(1)}(A) &  = {\beta} AF^{2}\,.
\end{align}
The Abelian Lagrangian is normalized by choosing {$\beta=3$}. In CS supergravity, {$\beta$ is} proportional to $\frac{1}{{\mathcal N}}-\frac{1}{4}$, so that {$\beta=0$} corresponds to CS supergravity invariant under the super AdS group  $SU(2,2|4)$. In that case, however, the theory has functionally dependent constraints around the most symmetric AdS background, that has to be specially dealt with \cite{Miskovic-Troncoso-Zanelli,canonical-CS-sectors,irregular}. The choice ${\beta}=3$ avoids this problem since it implies ${\mathcal N} \neq 4$. {The particular value 3 is chosen for simplicity of equations, as the constant always appears in the combination $\beta/3$.}

In CS supergravity there is a non-minimal coupling between geometry and the electromagnetic field brought about by the symmetric invariant tensor component $g_{1ab}$,
\begin{equation}
L_{\text{int}}=\frac{\alpha}{2}\,\left[ R^{ab}R_{ab}+\frac{2}{\ell^{2}}\, \left( R^{ab}e_{a}e_{b}-T^{a}T_{a}\right)  \right]  A\,,
\end{equation}
where $R^{ab}R_{ab}$ is the Lorentz Pontryagin four-form and $T^{a}T_{a}-R^{ab}e_{a}e_{b}=d\left( T^{a}e_{a}\right) $ is the Nieh-Yan invariant \cite{Nieh-Yan}. These define two topological invariants in four-dimensional Einstein-Cartan geometry, and the combination of both is the AdS Pontryagin four-form \cite{Chandia-Z}.

Varying the action (\ref{CS}) with respect to the connection $A^M$ yields the equations of motion $g_{MNK}\,F^N F^K=0$. More explicitly, they can be written as
\begin{align}
\delta e^a:\qquad 0 &  =\mathcal{L}_a =\frac{k}{4}\, \epsilon_{abcde}\,F^{bc}F^{de}-\frac{2\alpha}{\ell}\,T_{a}F\,,
\label{eqs without B 1}\\
\delta \omega^{ab}:\qquad 0 &  =\mathcal{L}_{ab}=\frac{k}{\ell}\, \epsilon_{abcde}\,F^{cd}T^{e}+2\alpha \,F_{ab}F\,,
\label{eqs without B 2}\\
\delta A:\qquad 0 &  =\mathcal{L}=FF+\frac{\alpha}{2}\, R^{ab}R_{ab} - \frac{\alpha}{\ell^2}\, d(T^a e_a ) \,.
\label{eqs without B 3}
\end{align}
These equations explicitly depend on the torsion tensor 2-form, $T^a=\frac{1}{2}\,T_{\mu\nu}^a\,dx^\mu dx^\nu$. If $T^a\neq0$, the manifold possesses both curvature and torsion, that describes a Rieman-Cartan spacetime.

In string theory, torsion $T_{\lambda\mu\nu}= e_{a\lambda}T_{\ \mu\nu}^{a}$ appears through the NS-NS field strength $H_{\lambda\mu\nu}=T_{\lambda\mu\nu}+T_{\mu\nu\lambda}+T_{\nu\lambda\mu}$ of the antisymmetric tensor field contained in the gravitation supermultiplet {\cite{Green-Schwarz-Witten}}. Then the $H$-torsion 3-form $H=T^{a}e_{a}$ is related to the completely antisymmetric part of the torsion tensor. Anomaly cancelation requires the inclusion of an AdS$\times U(1)$ CS terms, so that the Bianchi identity of the $H$-torsion takes the form
\begin{equation}
\frac{\alpha}{\ell^{2}}\,dH=FF+\frac{\alpha}{2}\,R^{ab}R_{ab}\,,
\end{equation}
which, in this case, is the dynamical equation (\ref{eqs without B 3}).

It is common in gravitation to use the second order formalism, where the fundamental fields ($e_{\mu}^{a},\omega_{\mu}^{ab}$) are replaced by the metric, $g_{\mu\nu}=\eta_{ab}\,e_{\mu}^{a}e_{\nu}^{b}$, and the affine connection $\Gamma_{\nu\mu}^{\lambda} =e_{a}^{\lambda}\left(  \partial_{\mu}e_{\nu}^{a} +\omega_{\mu}^{ab}e_{b\nu}\right)$ that defines parallel transport on the manifold $\mathcal{M}$. The symmetric part of the connection is the Christoffel symbol (determined by the metric), while its antisymmetric part is the torsion tensor, $T_{\ \mu\nu}^{\lambda} =\Gamma_{\nu\mu}^{\lambda}-\Gamma_{\mu\nu}^{\lambda}$. For more about the Riemann-Cartan spaces, see Appendix \ref{Conventions}.

The bosonic sector AdS$_5\times U(1)$  of CS supergravity action can be cast in the more familiar second order formalism with non-vanishing torsion. The purely gravitational part of the action includes the Gauss-Bonnet (GB) term and a negative cosmological
constant with fixed coupling constant $\ell^2/4$,
\begin{equation}
I_{\text{AdS}}\,=\frac{k}{\ell^{3}}\int d^{5}x\sqrt{-g}\,\left[  R + \frac{6}{\ell^{2}}+\frac{\ell^{2}}{4}\,\left(  R^{2} - 4R^{\mu\nu}R_{\nu\mu} + R^{\mu\nu\alpha\beta}R_{\alpha\beta\mu\nu}\right)  \right]  \,, \label{CS-AdS5}
\end{equation}
where the CS level $k=-\ell^{3}/16\pi G$ is related to the gravitational constant $G$. Note that in a spacetime with torsion, the curvature tensor $R_{\alpha \beta\mu\nu}$ is not symmetric under swapping of pairs of indices $[\alpha\beta]$ and $[\mu\nu]$, and the Ricci tensor $R_{\mu\nu}$ is not symmetric in $(\mu,\nu)$. This is because the connection also contains torsion-dependent terms. The choice of coupling constants in (\ref{CS-AdS5}) with ratios $6/\ell^2:1:\ell^2/4$ is such that the Lagrangian becomes a CS form \cite{Zanelli-CS-lectures}. For this unique ratio and in the absence of matter, the theory possesses a unique AdS vacuum. For a generic choice of coefficients, instead, the theory has two branches, each one having its own AdS$_5$ vacuum \cite{Boulware:1985wk}. As mentioned before, the uniqueness of the GB constant that maps GB to CS gravity also yields an enhancement of local symmetry from the Lorentz group, $SO(4,1)$, to the AdS$_5$ group, $SO(4,2)$, although it is hard to see the enhancement in this representation.

The electromagnetic kinetic term is described by the Abelian CS action,
\begin{equation}
I_{U(1)}=-\frac{1}{4}\int d^{5}x\,\epsilon^{\mu\nu\alpha\beta\lambda}F_{\mu\nu}F_{\alpha\beta}A_{\lambda}\,,
\end{equation}
and the interaction between gravity and the electromagnetic field explicitly involves the torsion tensor,
\begin{equation}
I_{\text{int}}=-\frac{\alpha}{8}\int d^{5}x\,\epsilon^{\mu\nu\alpha\beta\lambda}\left(  R_{\mu\nu\gamma\rho}\,R_{\ \ \alpha\beta}^{\gamma\rho}+\frac{4}{\ell^{2}}\,R_{\mu\nu\alpha\beta}-\frac{2}{\ell^{2}}\,T_{\ \mu\nu}^{\gamma} T_{\gamma\alpha\beta}\right)  A_{\lambda}\,.
\end{equation}
The field equations that extremize this action with respect to the metric are
\begin{equation}
R_{\mu\nu}-\frac{1}{2}\,g_{\mu\nu}\,R-\frac{3}{\ell^{2}}\,g_{\mu\nu}+{\cal H}_{\mu \nu}= \frac{\ell \alpha}{4k}\,\sqrt{-g}\,\epsilon_{\mu\alpha\beta\gamma\lambda}T_{\nu}^{\ \ \alpha\beta}F^{\gamma\lambda}\,, \label{grav}
\end{equation}
where the contribution of the quadratic terms in curvature is given by the Lanczos tensor,
\begin{align}
{\cal H}_{\mu\nu}  & =  \frac{1}{2}\,\left( R_{\mu\nu} R-2R_{\mu\alpha\nu\beta}R^{\beta\alpha}-2R_{\mu\alpha} R_{\ \nu}^\alpha+R_{\mu\lambda}^{\ \ \alpha\beta} R_{\alpha\beta\nu\lambda}\right) \nonumber\\
  &\quad -\frac{1}{8}\,g_{\mu\nu}\left( R^2-4 R^{\alpha\beta} R_{\beta\alpha}+R^{\alpha\beta\gamma\lambda} R_{\gamma\lambda\alpha\beta}\right)\,.
\end{align}
The electromagnetic field equations read
\begin{equation}
\epsilon^{\mu\alpha\beta\gamma\lambda}\left(  \frac{1}{4}\,F_{\alpha\beta}F_{\gamma\lambda}+ \frac{\alpha}{8} \,R_{\tau\sigma\alpha\beta}R_{\ \ \gamma\lambda}^{\tau\sigma} - \frac{\alpha}{2\ell^{2}}\,\nabla_{\alpha}
T_{\beta\gamma\lambda}\right) =0  \,,
\end{equation}
where $\nabla_{\alpha}$ is the covariant derivative defined with respect to the affine connection $\Gamma_{\beta\gamma}^{\alpha}$. The equations explicitly involving torsion are
\begin{align}
& 2\delta_{\lbrack\mu}^{\lambda}R_{\ \ \nu]\gamma}^{\alpha\beta} T_{\ \alpha \beta}^{\gamma}
- 4\delta_{\lbrack\mu}^{\lambda}R_{\ \nu]}^{\alpha}T_{\alpha} +  2\delta_{\lbrack\mu}^{\lambda}RT_{\nu]}+4\delta_{\lbrack\mu
|}^{\lambda}R_{\ \beta}^{\alpha}T_{\ \alpha|\nu]}^{\beta}\nonumber\\
&  +2R_{\ \ \mu\nu}^{\lambda\alpha}T_{\alpha}-4R_{\ [\mu}^{\lambda}T_{\nu
]}-4R_{\ \ [\mu|\beta}^{\lambda\alpha}T_{\ \alpha|\nu]}^{\beta}-2R_{\ \alpha
}^{\lambda}T_{\ \mu\nu}^{\alpha}\nonumber\\
&  +R_{\ \ \mu\nu}^{\alpha\beta}T_{\ \alpha\beta}^{\lambda}-4R_{\ [\mu
|}^{\alpha}T_{\ \alpha|\nu]}^{\lambda}+RT_{\ \mu\nu}^{\lambda}
+\frac{2}{\ell^{2}}\left(  2\delta_{\lbrack\mu}^{\lambda}T_{\nu]}
+T_{\ \mu\nu}^{\lambda}\right)  \nonumber\\
&  -\frac{\alpha\ell}{2k}\frac{1}{\sqrt{-g}}\,R_{\tau\sigma\mu\nu}F_{\alpha\beta
}\,\epsilon^{\lambda\tau\sigma\alpha\beta}-\frac{\alpha}{k\ell}\sqrt{-g}
\ g^{\lambda\alpha}\epsilon_{\alpha\mu\nu\tau\sigma}\,F^{\tau\sigma}=0 \,, \label{Torsion}
\end{align}
where $A_{[\mu}B_{\nu]}=\frac{1}{2}\left(  A_{\mu}B_{\nu}-A_{\nu}B_{\mu}\right)$ and $T_{\mu}=T_{\ \mu\alpha}^{\alpha}$. In our conventions, $\epsilon_{\mu\nu\alpha\beta\lambda}$ is the Levi-Civita tensor density, with $\epsilon_{01234}=1$, while $\frac{1}{\sqrt{-g}}\,\epsilon^{\mu\nu\alpha\beta\lambda}$ and $\sqrt{-g}\,\epsilon_{\mu\nu\alpha\beta\lambda}$ are covariantly constant tensors. Conventions for the $\epsilon$-symbol are given in Appendix \ref{Conventions}.

Although first order and tensorial formalisms are two alternative descriptions expected to give (at least classically) physically equivalent results, it is clear from the form of tensorial equations (\ref{grav})--(\ref{Torsion}) that they are too cumbersome to be useful. In contrast, first order formalism equations (\ref{eqs without B 1})--(\ref{eqs without B 3}) are simple, which justifies our working with the latter.

There is also a deeper reason to work with the vielbien and spin-connection instead of the metric and contorsion as fundamental fields. In the presence of fermions that live in the tangent space, or non-minimal couplings as in our case, the two formulations are not equivalent in general. A well-known example of a theory that does not possess (so far) first order formulation is New Massive Gravity; another example is Topologically Massive Gravity where the two formulations have different quantum anomalies. Thus, the fact that we work in the first order formalism is not just a simpler choice, but a necessity due to presence of torsional degrees of freedom, fermions and non-minimal interaction.

%%%%%%%%%%%%%%%%%%%%%%%%%%%%%%%%%%%%%%%%%%%%%
\section{Static, symmetric black holes   \label{Charged BH ansatz}} %  3  %
%%%%%%%%%%%%%%%%%%%%%%%%%%%%%%%%%%%%%%%%%%%%%
%%%%%%%%%%%%%%%%%%%
\subsection{The ansatz}  %  3.1  %
%%%%%%%%%%%%%%%%%%%

We are interested in finding an exact charged black hole solution to the field equations (\ref{eqs without B 1})--(\ref{eqs without B 3}). In the local coordinates $x^{\mu}=(t,r,x^{m})$ (with $m=2,3,4$), we seek black hole solutions with planar horizon, with a metric of the form
\begin{equation}
ds^2=-f^2(r)dt^2+\frac{dr^2}{f^2(r)}+r^2\delta_{mn}\,dx^m dx^n\,.\label{metric_ansatz}
\end{equation}
We restrict to spacetimes where the radial coordinate in non-negative. The generalization to the case of constant curvature horizons, ${\mathcal R}^{mn}_{kl}=\kappa \,\delta^{mn}_{kl}$ with $\kappa=0,\pm 1$, is straightforward. The only modification required is the shift in the metric function $f^2(r)\to f^2(r)+\kappa $. Since our motivation is in applications to holography, we restrict our analysis to the planar case $\kappa=0$.

For non-compact 3D space with the metric $\delta_{mn} dx^m dx^n$ and a specific form of the metric function $f(r)$, this solution represents a black 3-brane, while for discrete quotients of the 3D transverse space the geometry could be that of a topological black hole.

In the 3D transverse section, we use $i,j,k,\ldots=2,3,4$ to label tangent space indices, while the spacetime indices in a coordinate basis are labeled by $m,n,l,...=2,3,4$ referring to coordinates $(x^2, x^3, x^4):=(x,y,z)$. The third rank Levi-Civita tensor on the tangent to the transverse section is
\begin{equation}
^{( 3)}\epsilon_{mnl}:=\epsilon_{mnl}=\epsilon_{trmnl}\,,
\end{equation}
and $^{(3)}g_{mn}=\delta_{mn}$ is the corresponding flat metric. For more details on these conventions, see Appendix \ref{Conventions}.

Splitting the group indices as $a=(0,1,i)$, the vielbein can be chosen as
\begin{equation}
e^{0}=f(r)\,dt\,,\qquad e^{1}=\frac{dr}{f(r)}\,,\qquad e^{i}=r\,\delta_{m}^{i}\,dx^{m}:=r\,dx^{i}\,.
\end{equation}
The corresponding torsion-free spin connection, $\tilde{\omega}^{ab}$, and curvature $\tilde{R}^{ab}$, are given in Appendix \ref{BH ansatz}. In this ansatz, the torsion-free part of the Pontryagin form vanishes,
\begin{equation}
\tilde{R}^{ab}\tilde{R}_{ab}=0\, ,
\end{equation}
as it corresponds to a parity-even solution.

The isometry group of the five-dimensional AdS$_5$ black brane (\ref{metric_ansatz}) is $ISO(3) \times \mathbb{R}$ and is generated by seven Killing vectors: $\partial_{t}$ (time translation), $\epsilon_{mn}^{\ \ \ \ k}x^{m}\partial_{k}$, (rotations in the transverse section), and $\partial_{m}$ (translations in the three flat transverse directions). As explained in Appendix \ref{BH ansatz}, the gauge field 1-form $A$ compatible with these isometries has the form
\begin{equation}
A=A_{t}(r)\,dt+A_{r}(r)\,dr\,.  \label{la35}
\end{equation}

Let us assume that the space is torsion-free, $T^a=0$. In this ansatz, the component $\mathcal{L}_{0} \wedge dr=0$ of (\ref{eqs without B 1}) becomes
\begin{equation}
\left( f\,\frac{\partial f}{\partial r}-\frac{r}{\ell^2}\right)\left( f^2-\frac{r^2}{\ell^2}\right)=0\,,
\end{equation}
which leads to the uncharged black hole, $f^2(r)=\frac{r^2}{\ell^2}-\mu$ and the $U(1)$ field decouples from gravity. As shown next, the situation changes drastically if one assumes $T^{a}\neq0$. The torsion 2-form with the same isometries above, is given by the ansatz (see Appendix \ref{BH ansatz})
\begin{align}
T^{0}  &  =-\frac{\chi_{t}}{f}\,dt  dr\,,\qquad T^{1}   =f\chi_{r}\,dt  dr\,, \notag\\
T^{i}  &  =\frac{1}{r}\,\left(  \psi_{t}\,dt+\psi_{r}\,dr\right)
dx^{i}+\frac{\phi}{2r}\,\delta^{ik}\epsilon_{knm}\,dx^{n}  dx^{m}\,.
\label{T-ansatz}
\end{align}

The gravitational constant, $k=\ell^3/\ell^3_P$, where $\ell_P$ is the Planck length $\ell_P^3 = 16\pi G$, and the non-minimal coupling constant $\alpha$  are dimensionless, and the fields $A_\mu$ and $ \chi_\mu$ have units of inverse length, while $\psi_\mu$ and $\phi$ have dimensions of length and length square, respectively.

In the present ansatz, one can show that the {\it full} Pontryagin density need not vanish,
\begin{equation}
R^{ab} R_{ab}=d\left[ \frac{\phi}{r^4}\left( \frac{\phi^2}{12r^2}+ f^2 (\psi_r-r)^2 -\frac{\psi_t^2}{f^2}\right)\right]\epsilon_{knm}\,dx^k
dx^{n}  dx^{m}\,. \label{Pontryagin}
\end{equation}

Let us write now the field equations for this ansatz. In components, Eqs.(\ref{eqs without B 1}) become
\begin{align}
\mathcal{L}_{0}  &  =k\,\epsilon_{ijk}\,F^{1i}  F^{jk} +\frac{2\alpha}{\ell}\,T^{0}  F\,,\nonumber\\
\mathcal{L}_{1}  &  = - k\,\epsilon_{ijk}\,F^{0i}  F^{jk} - \frac{2\alpha}{\ell}\,T^{1}  F\,,\nonumber\\
\mathcal{L}_{i}  &  = k\,\epsilon_{ijk}\left(  F^{01}  F^{jk}-2F^{0j}  F^{1k}\right) - \frac{2\alpha}{\ell}\,T_{i}  F\,,
\end{align}
and Eqs.(\ref{eqs without B 2}) read
\begin{align}
\mathcal{L}_{01}  &  =\frac{k}{\ell}\, \epsilon_{ijk}\,F^{ij}  T^{k}-2\alpha\,F^{01}  F\,,\nonumber\\
\mathcal{L}_{0i}  &  =-\frac{k}{\ell}\,\epsilon_{ijk}\, \left(2F^{1j}  T^{k}+F^{jk}  T^{1}\right) -2\alpha\,F_{\ i}^{0}  F\,,\nonumber\\
\mathcal{L}_{1i}  &  =\frac{k}{\ell}\,\epsilon_{ijk}\,\left(2F^{0j}  T^{k}+F^{jk}  T^{0}\right) +2\alpha\,F_{\ i}^{1}  F\,,\nonumber\\
\mathcal{L}_{ij}  &  =\frac{2k}{\ell}\,\epsilon_{ijk}\,\left(F^{01}  T^{k}-F^{0k}  T^{1}+F^{1k}  T^{0}\right) +2\alpha\,F_{ij}  F\,.
\end{align}
All field equations are 4-forms so that their components are obtained by multiplication by 1-forms and using the identity $dt dr dx^m dx^n dx^k =-\epsilon^{mnk}\,d^5 x$.

In order to find the analytic solution it is convenient to write the equations of motion in components. Starting by equation $\mathcal{L}_{01}=0$, we find two nonvanishing components,
\begin{align}
0 &  =\left(-\frac{\psi_t^2}{f^2} + f^2 \left(\psi_r - r\right)^2 + \frac{\phi^2}{4r^2} - \frac{r^4}{\ell^2}\right)  \psi_t \,,\label{310bis} \\
0 &  =\left(-\frac{\psi_t^2}{f^2} + f^2 \left(\psi_r - r\right)^2 + \frac{\phi^2}{4r^2} - \frac{r^4}{\ell^2}\right)  \psi_r + \frac{\phi}{r} \left(\frac{r}{2}\phi' - \phi\right) \, ,
\end{align}
where the prime stands for the derivative with respect to $r$. Note that the interaction term proportional to $\alpha$ does not contribute to this particular field equations in this ansatz.  Assuming $\psi_t \psi_r \neq 0$, combining these two equations gives a differential equation in the field $\phi$ whose general solution is
\begin{equation}
\phi=2Cr^2\,, \label{C}
\end{equation}
with an integration constant $C$. The other equation implies that other fields must satisfy
\begin{equation}
\mathcal{T}(r) = -\frac{\psi_t^2}{f^2} + f^2 \left(\psi_r - r\right)^2 +C^2 r^2 - \frac{r^4}{\ell^2}=0\,.\label{Tau}
\end{equation}
Note that, without torsion ($\psi_p =0$, $C=0$), the only solution to $\mathcal{T}=0$ is AdS$_5$ with flat transverse section, $f^2 = \frac{r^2}{\ell^2}$, as expected.

Next, equation $\mathcal{L}_{0}=0$ yields two conditions,
\begin{align}
0 &  =\psi_{t}\left(  \chi_{t}-ff^{\prime}\right)  \mathcal{T}(r)\,,\nonumber\\
0 &  =\left(rf\,\psi_t \chi_r + f^2 f' \psi_r r - f^3 \psi_r + rf^3 \psi_r ' - r^2 f^2 f' + \frac{r^3}{\ell^2 }f \right) \mathcal{T}(r)\,, \label{la313}
\end{align}
which are identically satisfied for $\mathcal{T}(r)=0$.

Similarly, equation $\mathcal{L}=0$ in (\ref{eqs without B 3}), using $FF=0$, (\ref{Tinv}) and (\ref{Pontryagin}), can be written as
\begin{equation}
0=\left(  \frac{\phi}{r^{4}}\,\mathcal{T}(r)-\frac{\phi^{3}}{6r^{6}}\right)' , \label{la314}
\end{equation}
is also identically satisfied for $\mathcal{T}(r)=0$ and $\phi=2Cr^2$.

The non-vanishing components of equation $\mathcal{L}_{1}=0$ are also proportional to ${\mathcal T}(r)$,
\begin{align}
0 &  =dr \mathcal{L}_{1}\sim\left(  \chi_{t}\psi_{r}-r\chi_{t}%
-ff^{\prime}\psi_{r}+rff^{\prime}-\frac{r^{2}}{\ell^{2}}\right)
\,\mathcal{T}(r)\,, \\
0 &  =dt \mathcal{L}_{1}\sim\left(  -rf^{3}\chi_{r}\psi_{r}+r^{2}%
f^{3}\chi_{r}+f\psi_{t}+rf^{\prime}\psi_{t}-rf\psi_{t}^{\prime}\right)
\,\mathcal{T}(r)\,,  \label{la315}
\end{align}
and, again, they are not independent from Eq. (\ref{Tau}).

Let us focus first on solving ${\mathcal T}(r)=0$. Defining the new function $\eta(r)$ as
\begin{equation}
\psi_r :=r+\frac{\eta}{f}\,,
\end{equation}
Eq. (\ref{Tau}) reads
\begin{align}
\psi_t = \varepsilon_{\psi}\,f\sqrt{\eta^2 + C^2 r^2 -\frac{r^4}{\ell^2}} \,, \qquad \text{ }\varepsilon_{\psi}=\pm1\,, \label{psi}
\end{align}
where $\eta^2 + C^2 r^2 -\frac{r^4}{\ell^2} \geq 0 $, automatically solves (\ref{Tau})-(\ref{la315}).

Next, equation $\mathcal{L}_{i}=0$ reduces to
\begin{align}
\mathcal{E}(r) & = \frac{C\ell \alpha}{k}\,r^2 f F_{tr} - rf \chi_r \psi_t + f^2 \eta - rf^2 \eta' \nonumber\\
&  -f\,\frac{r^3}{\ell^2} + r^2 f \chi_t + r \chi_t \eta - r^2 f^2 f' - r f f' \eta = 0\,, \label{E(r)}
\end{align}
and, by the same token, $\mathcal{L}_{0i}=0$ and $\mathcal{L}_{1i}=0$ are automatically satisfied as well.

Finally, for $C\neq0$, equations $\mathcal{L}_{ij}=0$ lead to the last nontrivial expression, namely

\begin{equation}
\mathcal{S}(r)=\eta\chi_t - f\chi_r\psi_t - r^2 f \chi_t' + r^2 f f'^2 + r^2 f^2 f'' - \frac{r^2}{\ell^2}\,f=0\,. \label{la320}
\end{equation}

%%%%%%%%%%%%%%%%%%%%%%%%%%%%%
\subsection{Charged black hole solution} \label{Charged BH sol}   %  3 . 2   %
%%%%%%%%%%%%%%%%%%%%%%%%%%%%%
The field equations can now be solved to obtain explicit expressions for the fields $f(r)$, $A(r)$, $\phi(r)$, $\chi_p (r)$, $\psi_q(r)$, with $p,q=(r,t)$. The general solution to the system (\ref{310bis})-(\ref{la320}) is too cumbersome to extract physical information from it at first sight. It is better to begin by analyzing special cases; for instance, by studying solutions with only {\it some} non-zero components of the torsion.

Black hole solutions with non-vanishing torsion have been previously considered in the literature. For example, in Ref.\cite{Canfora:2007xs}, a solution with a metric of the form (\ref{metric_ansatz}) and axial torsion ($\phi (r) \neq 0$) was considered. That solution, however, is uncharged and so it does not require (and does not include) other components of the torsion ($\psi_p$ or $\chi_q$). In turn, the first example we would like to investigate is the simplest case in which, apart from $\phi(r)$, an additional component of the torsion is switched on, so that the resulting electric field is non-zero.

Consider, for example, the case with $\psi_t = \chi_r = \chi_t =0$, but with non-vanishing $\psi_r$ and $\phi $. In this case, the metric function $f(r)$ is given by
\begin{equation}
f^2(r)=\frac{r^2}{\ell^2}+br-\mu \,, \label{laf}
\end{equation}
where $b$ and $\mu $ are arbitrary constants.

The metric (\ref{metric_ansatz}) with (\ref{laf}) is the five-dimensional analogue of the {\it hairy} black hole solution considered in conformal gravity and massive gravity in three dimensions \cite{Hairy1,Hairy2}. This is also reminiscent of the solution of four-dimensional conformal gravity \cite{Riegert}, which also exhibits a linear damping off $\sim br$ in the metric function $f^2(r)$. In dimension grater than three, however, the metric is conformally flat only if $\mu =0$. Indeed, the components of the (torsionless) Weyl tensor of our five-dimensional solution read
\begin{eqnarray}
W^{0i} &=&\frac{\mu }{6r^2}\,e^0e^i,\qquad W^{1i}=-\frac{\mu }{6r^2}\,e^1 e^i\,, \notag \\
W^{01} &=&-\frac{\mu }{2r^2}\,e^0 e^1,\qquad W^{ij}=\frac{\mu }{6r^2}\,e^i e^j\,.
\end{eqnarray}
Thus, the parameter $b$ can be regarded as a gravitational hair. For some range of the parameters $\mu$ and $b$, the solution represents a topological black hole (or black brane). Indeed, these solutions have flat horizon and can be regarded as black branes in the case of non-compact base manifold with flat metric and $\mathbb{R}^3$ topology. For horizons of non-trivial topology, like $T^3$, or more general structure $\mathbb{R}^3/\Gamma $, where $\Gamma$ is a Fuchsian-like subgroup, these solutions represent topological black holes.

{If $b<0$, horizons exist provided $b^{2}\ell ^{2}+4\mu \geq 0$. These
horizons are located at
\begin{equation}
r_{\pm }=-\frac{b\ell ^{2}}{2}\left( 1\pm \sqrt{1+\frac{4\mu }{b^{2}\ell ^{2}
}}\right) .
\end{equation}
For $\mu = -\ell^2b^2/4$ and $b<0$ the solution is extremal in the sense that its two horizons coincide and the near horizon geometry is AdS$_2 \times \mathbb{R}^3$.}

{Notice that inner horizon $r_{-}$ is also positive if and only if $0>\mu
\geq -b^{2}\ell ^{2}/4$.
If $b>0$, instead, then the solution may only present one horizon, $r_{+}>0$, provided $\mu >0$. This horizon is located at
\begin{equation}
r_{+}=\frac{b\ell ^{2}}{2}\left( \sqrt{1+\frac{4\mu }{b^{2}\ell ^{2}}}-1\right) .
\end{equation}}

For $b\neq 0$, the solution (\ref{laf}) is asymptotically AdS$_5$ in a weaker sense. That is, the next-to-leading behavior of the metric components in the large $r$ limit is weaker than the standard asymptotically AdS conditions \cite{HT}. In particular, we find
\begin{equation}
g_{tt} \sim \frac{r^2}{\ell ^2} + {\mathcal O}(r)\,, \qquad  g_{rr} \sim \frac{\ell ^2}{r^2} + {\mathcal O}(1/r^3)\,. \label{327}
\end{equation}
Notice that the ${\mathcal O}(r)$ term can be absorbed by the change $r = r' - b \ell^2 /2$, so that a metric obeying asymptotic behavior (\ref{327}) can be turn into one obeying the standard (stronger) asymptotic behavior
\begin{equation}
g_{tt} \sim \frac{r^2}{\ell ^2} + {\mathcal O}(1)\,, \qquad  g_{rr} \sim \frac{\ell ^2}{r^2} + {\mathcal O}(1/r^4)\,. \label{327bis}
\end{equation}
However, being a $b$-dependent coordinate transformation, the shift $r = r' - b \ell^2 /2$ is not enough to change a whole set of metrics obeying (\ref{327}) into a set of metrics obeying (\ref{327bis}), but merely in making $b$ to dissapear from the leading piece of the large $r$ behavior of a particular member of such a set of metrics. This remark is important because, in the context of holography, the notion of the set of asymptotically AdS solutions \cite{HT} is the one that becomes relevant. It is also worth pointing out that such shift in the coordinate $r$ does not suffice to eliminate the parameter $b$ completely from the metric, but only from its leading terms in the large $r$ behavior. In fact, the parameter $b$ represents an actual parameter of the solution, just as $\mu $, and can not be eliminated by a coordinate transformation. This can be verified by explicitly computing the scalar curvature associated to metric (\ref{metric_ansatz}) with (\ref{laf}), which reads
\begin{equation}
R=-\frac{20}{\ell^2}+\frac{12b}{r}+\frac{6\mu}{r^2},
\end{equation}
and explicitly depends both on $\mu$ and on $b$. Nevertheless, the fact that the shift $r = r' - b \ell^2 /2$ makes the $g_{tt}$ component of the metric to take the form in (\ref{327bis}) leads us to argue that the physical mass of the solution would be given as a function of the the linear combination $\mu' = \mu + (b \ell/2)^2$ and not just $\mu $.

The axial component of the torsion remains $\phi(r)=2C r^2$, with $C$ a third independent integration constant. The new non-vanishing component of the torsion is now
\begin{equation}
\psi_r = r\frac{\sqrt{r^2+\ell^2br-\ell^2\mu } -\varepsilon_\psi \sqrt{r^2-\ell^2C^2} }{\sqrt{r^2+\ell^2br-\ell^2\mu }}.
\end{equation}
with $\varepsilon_\psi= \pm 1$; we consider the case $\varepsilon_{\psi}=+1$. Recall that the other components are $\psi_t=\chi_r=\chi_t=0$.

From the field equations one easily verifies that for this configuration the electric field is non-zero and for $\varepsilon_\psi=1$ it is given by
\begin{equation}
A_t = \Phi -\frac{k}{C\ell \alpha} \left[\frac{r^2}{\ell^2}+\frac{br}{2}-\sqrt{\left(\frac{r^2}{\ell^2}
+br-\mu \right)\left(\frac{r^2}{\ell^2}-C^2\right)}\right],\quad A_r =0\,, \label{Adiv}
\end{equation}
where $\Phi $ is a new arbitrary constant.

At large $r$, the electrostatic potential $(\ref{Adiv})$ goes as
\begin{equation}
A_t (r) \sim Const+\frac{k\ell}{4C\alpha}\,\left(\mu-C^2+\frac{b^2\ell^2}{4}\right)\,\frac{b}{r}+{\mathcal O}(1/r^2)\,.
\label{asymptotic At}
\end{equation}

This means that, for $b\neq 0$ and $b\neq \pm 2\sqrt{C^2-\mu}/\ell$, the field strength $F=dA$ behaves asymptotically like $F_{rt}\sim {\mathcal O}(1/r^2)$, and this implies that the solution exhibits infrared divergent field energy, and is in this sense reminiscent of the self-gravitating Yang monopole solutions \cite{GibbonsTownsend}. On the other hand, $A_t(r)$ remains finite for $0\leq r \leq \infty$. The curve $\ell b = \pm 2\sqrt{C^2-\mu}$ in the parameter space seems special. In particular, this curve includes the point $b=\mu -C^2=0$ with $\varepsilon_\psi=1$, at which the asymptotic electric field loses the $1/r$ term in the expansion (\ref{asymptotic At}) and the field energy becomes finite. In fact, at this point the electric field vanishes ($A_t = Const.$) and the solution (\ref{laf})--(\ref{Adiv}) reduces to $\psi_r=0$, $\phi(r)=2Cr^2$, with $f^2(r)=r^2/\ell^2 -C^2$, which turns out to be a special case of the solution found in Ref. \cite{Canfora:2007xs}. In the next section we discuss the relation with that solution in more detail.

{On the curve $\ell b = \pm 2\sqrt{C^2-\mu}$ the mass of the solution can be seen to give
\begin{equation}
M=\frac{3\ell^2\,\text{Vol}(\gamma_3)}{16\pi G }\,\left(\mu +\frac{b^2\ell^2}{4}\right)^2\,, \label{MASS}
\end{equation}
where $\text{Vol}(\gamma_3)$ stands for the volume of the horizon three-surface.
This value for the mass can be computed by the Hamiltonian method \cite{Regge-Teitelboim}, see Appendix \ref{Mass RT}.
Notice that expression (\ref{MASS}) is positive definite provided horizons exist,
and it vanishes at the extremal case $r_{+}=r_{-}=\ell ^{2}b/2$.}

The Hawking temperature of black branes solutions (\ref{laf}) is given by
\begin{equation}
T= \frac{1}{4\pi \ell^2}\, (r_+ - r_-)\, ,
\end{equation}
which also vanishes when $\mu = -\ell^2b^2/4$, namely when $r_+ = r_- = - \ell^2b/2$.

{On the other hand, an entropy formula for these solutions can be inferred
from assuming the the first law of black holes thermodynamics actually holds.
In fact, assuming $\delta M=T\,\delta S$, the entropy would take the form%
\begin{equation}
S=\frac{(r_{+}-r_{-})^{3}\,\text{Vol}(\gamma_3)}{16G}.  \label{ENTROPY}
\end{equation}}

{As probably expected, expression (\ref{ENTROPY}) scales as $\sim r_{+}^{3}/G$
in the limit $r_{+}>>r_{-}$, reproducing the standard behavior of $b=0$
topological black holes of locally flat horizons in five-dimensional
Chern-Simons gravity. In general, being solutions of a higher-curvature
theory, Chern-Simons black holes do not obeyed the area law. In particular,
we see in (\ref{ENTROPY}) that for these solutions the entropy goes as the
cube of the distance between the two horizons multiplied by the volume of
the $r$-constant surfaces, $\text{Vol}(\gamma_3)$.}

%%%%%%%%%%%%%%%%%%%%%%%%%%
\subsection{Torsion and degeneracy}     %  3 . 3   %
%%%%%%%%%%%%%%%%%%%%%%%%%%
Let us now consider the cases in which other components of the torsion are switched on. The next example is that with non-vanishing $\psi _t$. In that case one gets
\begin{equation}
\begin{array}
[c]{llll}
f^2 & =\frac{r^2}{\ell^2} + br - \mu + \theta\,, & \phi & = 2Cr^2 \,,\\
A_t & =\Phi - \frac{k}{C\ell \alpha}\left(rff' + \frac{f\eta}{r} \right)\,, & \psi_r & =  r + \frac{\eta}{f}  \,,\\
\chi_r & =\frac{r^2 \theta''}{2\psi_t }\,, & \psi_t & = \varepsilon_\psi f\,\sqrt{\eta^2 + C^2 r^2 - \frac{r^4}{\ell^2}}\,,
\end{array}
\end{equation}
where $\theta(r)$ is an arbitrary function. Here, a distinctive feature of Chern-Simons (super)gravity theories is found; that is, the appearance of arbitrary functions that arise from degeneracies in the symplectic structure on certain special submanifolds of phase space. At those degeneracy surfaces the system acquires extra gauge symmetry and looses dynamical degrees of freedom. This is a generic feature of higher dimensional CS systems \cite{Banados-Garay-Henneaux,canonical-CS-sectors}, but it has been known to exist in all generic Lovelock theories \cite{Wheeler:1985qd,Teitelboim-Z} (see also the discussions in \cite{TroncosoOliva1, TroncosoOliva2, Oliva3} and references therein), as well as in many mechanical systems \cite{Saavedra-Troncoso-Z}. In the above solution, both $\chi_r (r)$ and $\psi_r (r)$ remain undetermined, as $\theta (r)$ and $\eta (r)$ are arbitrary functions of $r$. General Lovelock theory has a pathological structure of its phase space because of the non-invertible relation between the metric and its conjugate momentum \cite{Teitelboim-Z}. This introduces an indeterminacy in the dynamical evolution and leads to degenerate dynamics. At the CS point of the parameter space, the degeneracy is much more dramatic and of a peculiar class, generically yielding a plethora of undetermined free functions.

This phenomenon occurs also in the present case for $\psi_{t}=0$ and $\chi_{t}\neq0$. Then we have $\eta (r)=\varepsilon_{\eta}
\,r\sqrt{{r^{2}}/{\ell^{2}}-C^{2}}\geq0$ (the manifold is not complete),  there is also one arbitrary function $\theta(r)$ and the fields read
\begin{equation}
\begin{array}
[c]{llll}
f^{2} & =\frac{r^{2}}{\ell^{2}}+br-\mu+\theta\,, & \phi & =2Cr^{2}\,,\\
A_{t} & =\Phi-\frac{k}{C\ell \alpha}\left(  rff^{\prime}+\frac{f\eta}{r}
-r\chi_{t}\right)  \,, & \psi_{r} & =  r+\frac{\eta}{f}  \,,\\
\chi_{t} & =\chi_{t}^{0}\,\exp\left(  \varepsilon_{\eta}\varepsilon_{f}
{\displaystyle\int}
dr\,\frac{\sqrt{\frac{r^{2}}{\ell^{2}}-C^{2}}}{r^{2}\sqrt{\frac{r^{2}}
{\ell^{2}}+br-\mu+\theta}}\right)  +\bar{\chi}_{t}\,, &  &
\end{array}
\end{equation}
where $\chi_{t}^{0}$ is a constant and $\bar\chi_t(r)$ is a partial solution of the non-linear differential equation
\begin{equation}
\bar\chi_t' - \varepsilon_\eta \varepsilon_f \, \frac{\sqrt{\frac{r^2}{\ell^2}-C^2}}{r^2\sqrt{\frac{r^2}{\ell^2} + br - \mu + \theta}}\,\bar\chi_t = \frac{\theta''}{2}\,.
\end{equation}
In the case of more general solutions (e.g. $\chi_t\neq 0$), the number of arbitrary functions increases, as will shown below.

%%%%%%%%%%%%%%%%%%%%%%
\subsection{General solution} \label{General sol} %  3 .  4  %
%%%%%%%%%%%%%%%%%%%%%%

Consider now the general solution within the proposed form (\ref{metric_ansatz}), (\ref{la35})-(\ref{T-ansatz}). The spherically symmetric ansatz depends on eight independent functions, namely $f$, $A_{t}$, $A_{r}$, $\psi_{t}$, $\psi_{r}$, $\chi_{t}$, $\chi_{r}$, and $\phi$. In the static case, the component $A_r$ does not change the electric field $F_{tr}=-A_t'$ and can be gauged away to $A_r=0$, therefore we take
\begin{align}
A_{r}  &  =0\,,\quad \phi = 2Cr^2 \,,\quad \psi_r = r + \frac{\eta}{f}\,, \quad \psi_t = \varepsilon_\psi f\,\sqrt{\eta^2 + C^2 r^2 - \frac{r^4}{\ell^2 }}\,, \\
& \qquad \eta^2 + C^2 r^2 - \frac{r^4}{\ell^2} \geq 0\,, \qquad \varepsilon_\psi =\pm 1\,. \label{eta(r)}
\end{align}

The metric function $f(r)$ can be determined from $\mathcal{S}(r)=0$,
\begin{equation}
\frac{\eta\chi_t}{rf} - \frac{\chi_r \psi_t}{r} + r \left(ff' - \frac{r}{\ell^2} - \chi_t \right)'=0\,, \label{f(r)}
\end{equation}
while the electric potential $A_{t}$ is calculated from $\mathcal{E}(r)=0$,
\begin{align}
A_t & = \frac{k}{C\ell \alpha}\int\frac{dr}{r}\left[-rff' - \frac{r^2}{\ell^2} - (f\eta)' + \frac{f\eta}{r} - \chi_r \psi_t + \left(r + \frac{\eta}{f} \right)\chi_t \right]  + \Phi \nonumber\\
&  =\frac{k}{C\ell \alpha}\left[-\frac{f^2}{2} - \frac{r^2}{2\ell^2} - \frac{f\eta}{r} + \int dr\left(\frac{\eta\chi_t}{rf}-\frac{\chi_r \psi_t}{r} + \chi_t \right)\right] + \Phi\,,
\end{align}
where, again, $\Phi $ is an arbitrary constant. Integrating by parts in Eq.(\ref{f(r)}) yields
\begin{equation}
\int dr\left(\frac{\eta\chi_t}{rf} - \frac{\chi_r \psi_t}{r} - \chi_t \right) = - rff' + r\chi_t + \frac{f^2}{2}+ \frac{r^2}{2\ell^2}\,.
\end{equation}
Plugging this integral into the expression for $A_{t}$, the electric potential is obtained as
\begin{equation}
A_t = \Phi - \frac{k}{C\ell \alpha} \left(rff' + \frac{f\eta}{r} - r \chi_t \right)  \,.
\end{equation}
Note that this expression for $A_{t}$\ suggests that this solution is non perturbative in the sense that it has a dependence $1/\alpha$. However, it is possible to rescale $A_{t}\rightarrow \alpha A_{t}$ as with the electric charge in the Maxwell field. Notice that the axial torsion $C\neq 0$ also enters in the solution in a seemingly non-perturbative way.

Finally, we can write equation for $f(r)$ given by (\ref{f(r)}) as follows
\begin{equation}
\left(ff' - \frac{r}{\ell^2} - \chi_t \right)' = \frac{\chi_r \psi_t}{r^2} - \frac{\eta\chi_t}{r^2 f}\,. \label{fprime}
\end{equation}

The arbitrary functions $\eta(r)$, $\chi_{t}(r)$ and $\chi_{r}(r)$ can  be replaced a different set of arbitrary functions $\theta(r)$,
$\theta_{t}(r)$ and $\theta_{r}(r)$ which we choose as follows
\begin{eqnarray}
\eta     &=& \frac{r^2 f\,\theta_t'}{\theta' - \theta_r' + \theta_t} \,, \label{eta} \\
\chi_t &=& \frac{r^2 f}{2\eta\,\theta_t'} \,,  \label{chit} \\
\chi_r &=& \frac{r^2 \theta_r''}{2\psi_t}\, . \label{chir}
\end{eqnarray}

The transformation $(\chi_t, \chi_r, \eta) \rightarrow (\theta_t, \theta_r, \theta)$ is invertible given $\psi_t $, $\chi_t, \eta \neq 0$, provided  $\eta$ satisfies (\ref{eta(r)}), or equivalently
\[
\theta' - \theta_r ' + \theta_t \neq 0\,,\qquad \left(\frac{rf\,\theta_t '}{\theta' - \theta_r ' + \theta_t}\right)^2 > \frac{r^2}{\ell^2} - C^2.
\]
Equations (\ref{chit}) and (\ref{chir}) can be integrated directly as
\begin{equation}
\theta_t(r) =\int\limits^r ds\, \frac{2\chi_t(s) \eta(s)}{s^2 f(s)} \,, \quad \theta_r' (r) = \int\limits^r ds\, \frac{2\chi_r(s) \psi_t(s)}{s^2 } \, .
\end{equation}
Combining these with (\ref{fprime}), (\ref{eta}) can be integrated for $\theta$ as function of $f$, to finally give
\begin{equation}
f^{2}=\frac{r^{2}}{\ell^{2}}+b\,r-\mu+\theta(r)\, . \label{la327}
\end{equation}
Since $\theta(r)$ is arbitrary, it can absorb all $r$-dependent terms, including constants $b$ and $\mu $. This would, however, change the behavior of other fields that depend on $\theta$, so that we prefer to keep the form (\ref{la327}) for notational convenience.

In terms of the functions $\theta$s, the metric and electromagnetic fields read
\begin{eqnarray}
f^2 &=&\frac{r^2}{\ell ^2} + br - \mu +\theta \,,  \nonumber \\
A_t &=&\Phi - \frac{k}{C\ell \alpha}\left[ \frac{r^2}{\ell ^2} + \frac{br}{2} + \frac{r\theta_t '}{\theta' - \theta_r '+ \theta_t}
\left(\frac{r^2}{\ell^2} + br - \mu +\theta \right) + \frac{r (\theta_r' - \theta_t) }{2}\right] \,, \nonumber \\
A_{r} &=&0\,, \label{General1}
\end{eqnarray}
while the components of torsion are
\begin{eqnarray}
\phi  &=& 2Cr^2\,,  \nonumber \\
\psi _t &=& \varepsilon _\psi\varepsilon_f\,r\sqrt{\frac{r^2}{\ell^2}+br-\mu
+\theta}\sqrt{\left( \frac{r^2}{\ell^2}+br-\mu +\theta\right) \left( \frac{r\,\theta_t^\prime}{\theta^\prime
-\theta_r^\prime+\theta _t}\right)^2+C^2-\frac{r^2}{\ell^2}}\,,\nonumber \\
\psi _r &=& r\left( 1+\frac{r\,\theta _{t}^{\prime }}{\theta ^{\prime
}-\theta _{r}^{\prime }+\theta _{t}}\right) \,,  \nonumber \\
\chi _{t} &=& \frac{\theta ^{\prime }-\theta _{r}^{\prime }+\theta _{t}}{2}\,,\nonumber \\
\chi _{r} &=&\frac{\varepsilon _{\psi }\varepsilon _{f}\,r\,
\theta_{r}^{\prime \prime }}{2\,\sqrt{\frac{r^{2}}{\ell ^{2}}+br-\mu +\theta }
\sqrt{\left( \frac{r^{2}}{\ell ^{2}}+br-\mu +\theta \right)
\left( \frac{r\,\theta _{t}^{\prime }}{\theta ^{\prime }
-\theta _{r}^{\prime }+\theta _{t}}\right) ^{2}+C^{2}-\frac{r^{2}}{\ell ^{2}}}}\,,
\label{General2}
\end{eqnarray}
and we observe a high degree of degeneracy, brought about by the arbitrariness in $\theta _t (r),\theta _r (r)$, and $\theta (r)$.

As we said before, the appearance of {arbitrary functions is a distinctive feature of CS gravities, although it is not an exclusive property of the Chern-Simons form, nor is it due to the} presence of torsion. Indeed, already Wheeler noticed that so-called ``geometrically free solutions'', whose metric is not fully determined by field equations, typically appear in Lovelock gravity\footnote{Again, we emphasize that, apart from the indeterminacy that higher-curvature Lovelock theory has {\it per se}, the CS theory corresponds to a peculiar point of the parameter space at which the degeneracy drastically increases due to the symmetry enhancement.} when its coupling constants are such that it has a degenerate AdS vacuum \cite{Wheeler:1985qd}. In this sense, CS AdS theory is a special  Lovelock gravity in odd dimensions whose vacuum has maximal possible degeneracy.

On the other hand, metrics with undetermined components were reported in higher-dimensional theories in the torsionless case as well, e.g., in Einstein-Gauss-Bonnet AdS gravity when the transverse section of the metric is maximally symmetric \cite{Zegers:2005vx}. If the metric functions are time dependent they can still possess undetermined components in Chern-Simons theories \cite{Oliva3}.

It has been argued that the arbitrariness in the metric that appear in five-dimensional CS AdS gravity can be removed either by gauge-fixing \cite{Aros:2006qc}, or by changing the cosmological constant so that CS gravity becomes effectively EGB gravity \cite{Banados:2001hm}. The solution of Ref.\cite{Aros:2006qc}, however, is still degenerate even though the gauge-fixing hides the original arbitrariness in the metric.

In Section 6 we examine the canonical structure of CS AdS gravity about the sector of solutions of interest here in order to understand better the origin of these arbitrary functions.

%%%%%%%%%%%%%%%%%%%%%%%%%%%%%%%%%%%%%%%%%%%
\section{Comparison with the axial-torsion solution}  \label{Comparison}    %   4   %
%%%%%%%%%%%%%%%%%%%%%%%%%%%%%%%%%%%%%%%%%%%
Before going into the analysis of the peculiarities of the sector of the solution space we are considering, it is of particular importance to compare it with, at first sight, a very similar uncharged black hole geometry presented in Ref.\cite{Canfora:2007xs} that possesses only the axial component of torsion. As  mentioned before, the axial-torsion solution and the one presented in Sect. 3 coincide at a particular point of the space of solutions. More precisely, a special case of our solution (\ref{laf})-(\ref{Adiv}) coincides with the axial-torsion one (see Eqs. (19)-(20) in \cite{Canfora:2007xs} and/or Eqs. (\ref{U1})-(\ref{U1}) below). Then, a natural question is whether the whole family of axial-torsion solutions actually corresponds to a particular case of ours for $\psi_t=\psi_r=\chi_r=\chi_t=0$ and constant $A_t$.

As we shall see below, the answer is no. In fact, the two solutions belong to different branches of the space of solutions and they only meet at a particular point of their respective parameter spaces. Roughly speaking, while the solution considered here amounts to solve ${\mathcal T}(r)=0$ so that it possesses five non-vanishing torsion components (see for instance (\ref{Tau})), the axial-torsion solution in general solves equations of the form $\chi_p {\mathcal T}(r)=0$ and $\psi_p {\mathcal T}(r)=0$, with $p=r,t$, by choosing $\psi_p = \chi_p=0$. Both solutions (five-component torsion and axial-torsion ones) coincide at the point $\chi_p{\mathcal T}(r)=\psi_p{\mathcal T}(r)={\mathcal T}(r)=0$, which occurs for $b=0$ and $\mu=C^2$.

More concretely, the axial-torsion solution has the form
\begin{equation}
\chi_p=0\,,\qquad\psi_p=0\,,\qquad\phi=2Cr^{2}\,. \label{U1}
\end{equation}
The metric, on the other hand, is given by (\ref{metric_ansatz}) with $f$ completely determined to be
\begin{equation}
f^{2}(r)=\frac{r^{2}}{\ell^{2}}-\mu\,. \label{U2}
\end{equation}
The indeterminacy in the metric can be removed by imposing all components of the torsion except the axial one ($\phi(r)$), to vanish. Naively, this choice resembles fixing of the functions $\theta$; however, we will show that this  corresponds to a new branch of solutions independent from ours.

It can be explicitly shown that these two solutions are not connected by a gauge transformations. Let us denote by $\mathbf{A}$ our solution (\ref{General1}) and (\ref{General2}) for the symmetric ansatz of the theory when all five torsional components $\phi$, $\psi_t$, $\psi_r$, $\chi_t$, and $\chi_r$ are switched on; and let us denote by $\mathbf{\bar{A}}$ the axial-torsion solution (\ref{U1}) and (\ref{U2}). In the latter case,
\begin{equation}
\begin{array}
[c]{llll}
\bar{T}^{0} & =0\,, & \bar{F}^{01} & =-\left(  \bar{f}\bar{f}^{\prime}\right)
^{\prime}\,dt  dr\,,\medskip\\
\bar{T}^{1} & =0\,, & \bar{F}^{0i} & =-\bar{f}^{2}\bar{f}^{\prime}\,dt
dx^{i}\,,\medskip\\
\bar{T}^{i} & =\frac{\bar{\phi}}{2r}\,\delta^{ik}\epsilon_{knm}\,dx^{n}%
dx^{m}\,, & \bar{F}^{1i} & =-\bar{f}^{\prime}dr  dx^{i}-\frac{\bar
{f}\bar{\phi}}{2r^{2}}\,\epsilon_{\ jk}^{i}\,dx^{j}  dx^{k}\,,\medskip\\
&  & \bar{F}^{ij} & =-\left(  \bar{f}^{2}+\frac{\bar{\phi}^{2}}{4r^{4}%
}\right)  \,dx^{i}  dx^{j}\,.
\end{array}
\end{equation}

We are interested in finding a gauge transformation $g\in SO(4,2)\times U(1)$, if it exists, that maps $\mathbf{\bar{A}}$ into $\mathbf{A}$  according to the transformation law
\begin{equation}
\mathbf{F}=g^{-1}\mathbf{\bar{F}}\,g\,.
\end{equation}
%The same question can be asked for diffeomporphisms: Is there a $\xi=\xi^{\mu
%}\partial_{\mu}$ that maps $\mathbf{\bar{A}}$ to $\mathbf{A}=\mathbf{\bar
%{A}+\pounds }_{\xi}\mathbf{\bar{A}}$?

Consider first the infinitesimal gauge transformation, $g=e^{\mathbf{\Lambda}}\simeq1+\mathbf{\Lambda}$. The solution $\mathbf{T}$ of the form (\ref{T-ansatz}) is connected to the axial-torsion solution $\mathbf{\bar{T}}$ with non-trivial fields $\bar{\phi}=2\bar{C}r^{2}$ and $\bar{f}^{2}={r^{2}}/{\ell^{2}}-\bar{\mu}$ if there exists a $\mathbf{\Lambda}$ such that
\begin{equation}
\mathbf{T}=\mathbf{\bar{T}}+\delta_{\Lambda}\mathbf{\bar{T}\,,}
\end{equation}
where, in components
\begin{equation}
\delta_{\Lambda}T^{a}=R^{ab}\varepsilon_{b}-\lambda^{ab}T_{b}+\frac{1}{\ell^{2}}\,\varepsilon^{b}e^{a}  e_{b}\,.
\end{equation}
The transformation law of the gauge fields in components is given by Eq.(\ref{gauge tr}).

Let us start with $T^i=\bar{T}^{i}+\delta_{\Lambda}\bar{T}^{i}$, that is,
\begin{align}
T^{i} &  =\bar{T}^{i}+\left(  \bar{f}\bar{f}^{\prime}-\frac{r}{\ell^{2}}\right)
\left(  -\bar{f}\,\varepsilon^{0}dt+\frac{\varepsilon^{1}}{\bar{f}}\,dr\right) dx^{i}\nonumber\\
&  -\left(  \bar{f}^{2}-\frac{r^{2}}{\ell^{2}}+\frac{\bar{\phi}^{2}}{4r^{4}}\right)
\delta_{n}^{i}\varepsilon_{m}\,dx^{n}  dx^{m}\nonumber\\
&  +\left(  \frac{\bar{f}}{r}\,\varepsilon^{1}\delta^{ik}-
\lambda^{ik}\right)  \frac{\bar{\phi}}{2r}\,\epsilon_{knm}\,dx^{n}  dx^{m}\,.
\end{align}
Then the component along $dt  dr$ is trivially satisfied, and the components along $dt  dx^{m}$ and $dr  dx^{m}$ yield
\begin{align}
\frac{1}{r}\,\psi_{t} &  =-\left(  \bar{f}\bar{f}^{\prime}-\frac{r}{\ell^{2}%
}\right)  \,\bar{f}\varepsilon^{0}\,,\nonumber\\
\frac{1}{r}\,\psi_{r} &  =\left(  \bar{f}\bar{f}^{\prime}-\frac{r}{\ell^{2}%
}\right)  \frac{\varepsilon^{1}}{\bar{f}}\,.
\end{align}
For the axial-torsion solution, we have $\bar{f}\bar{f}'-{r}/{\ell^{2}}=0$, and the parameters $\varepsilon^0$ and $\varepsilon^1$
in the first two equations cannot be solved, thus the components $\psi_{t}$ and $\psi_{r}$ cannot be switched on by applying such a gauge transformation.

The equation along $dx^{n}  dx^{m}$, multiplied by $\epsilon^{jnm}$, fixes the following gauge parameters,
\begin{equation}
\varepsilon^{1}=\frac{C-\bar{C}}{\bar{C}}\frac{r}{\bar{f}}\,,\qquad \lambda^{ij}=\epsilon^{ijm}\varepsilon_{m}\,\frac{\bar{C}^{2}-\bar{\mu}}{2\bar{C}r}\,.
\end{equation}
Similarly, the equations for $T^{0}$ and $T^{1}$ imply that the components $\chi_{t}$ and $\chi_{r}$ cannot be switched on by gauge transformations, and also
\begin{equation}
\lambda^{0i}=0\,,\qquad\lambda^{1i}=-\frac{\bar{f}}{r}\,\varepsilon^{i}\,.
\end{equation}
This gauge transformation does not introduce new components, but merely changes the values of the integration constants of already existing fields. A similar conclusion is reached for the diffeomorphisms as well -- they just map one integration constant to another, and cannot switch off (or on) the torsion components.

Then, in spite of their similarities, the two solutions belong to different branches. Nevertheless, and as we already discussed, there is a limit in which our solution coincides with the axial-torsion one. In fact, asking that the non-axial torsion vanish ($\psi_p,\chi_p=0$) and $f^{2}=\frac{r^{2}}{\ell^{2}}-\mu$, we get $\eta=-rf$ and, as a consequence, the electromagnetic field vanishes, $A_{t}=\Phi-\frac{k\mu}{C\ell \alpha} =Const$. Thus, this limit is possible only for fixed values of the coupling constants $\mu=C^{2}$, so the axial-torsion solutions with $\mu\neq C^2$ are not accessible from our solution space.

Another way of seeing that both solutions belong to different branches is by direct analysis of the field equations. When the torsional degrees of freedom $\psi_p,\chi_p$ vanish, it is possible to solve the equations so that
\begin{equation}
\mathcal{T}(r)=r^{2}\left(  f^{2}+C^{2}-\frac{r^{2}}{\ell^{2}}\right) \neq0\,.
\end{equation}
Comparing with (\ref{Tau}) it is clear that the two solutions would coincide only for the special case $C^{2}=\mu$.

Yet another way to verify that the solutions belong to gauge-inequivalent sectors is by showing that they have different Casimir invariants. Namely, if $\mathbf{\bar{F}}$ and $\mathbf{F}$ are not connected by any finite gauge transformation $g$, they will have different gauge invariants of $SO(4,2)\times U(1)$ such as, for example, $U(1)$ and AdS invariants
\begin{align}
F_{U(1)}^2 &  =F_{\mu\nu}F^{\mu\nu}\,,\nonumber\\
F_{\text{AdS}}^2 &  =g^{\mu\alpha}g^{\nu\beta}\,F_{\mu\nu}^{AB}\,
F_{\alpha\beta}^{CD}\,\eta_{AC}\eta_{BD}=F_{\ \ \mu\nu}^{ab}F_{ab}^{\ \ \ \mu\nu}
-\frac{2}{\ell^{2}}\,T_{\ \mu\nu}^{a}\,T_{a}^{\ \,\mu\nu}\,,
\end{align}
where the subscript AdS refers to the $SO(4,2)$ piece of the group.

We shall choose the constant electric potential so that $F_{U(1)}^2=0$,
\begin{equation}
A_{t}=\Phi-\frac{k}{C\ell \alpha}\,\beta\,,
\end{equation}
where $\beta=Const$. This condition determines the torsion component as
\begin{equation}
\chi_{t}=ff^{\prime}+\frac{f\eta}{r^{2}}-\frac{\beta}{r}\,.
\end{equation}
Furthermore, non-vanishing components of $F^{AB}$ of the axial-torsion solution are
\begin{align}
T^{i} &  =Cr\,\epsilon_{\ mn}^{i}\,dx^{m}  dx^{n}\,,\nonumber\\
F^{ij} &  =\left(  \mu-C^{2}\right)  \,\delta_{m}^{i}\delta_{n}^{j}
\,dx^{m}  dx^{n}\,,\nonumber\\
F^{1i} &  =-Cf\,\epsilon_{\ mn}^{i}\,dx^{m}  dx^{n}\,,
\end{align}
leading to the AdS Casimir  invariant in the form
\begin{equation}
 F_{\text{AdS}}^2=\frac{12}{r^{4}}\,\left(  \mu^{2}
+C^{4}-6\mu C^{2}\right)  \,.\label{CGT invariant}
\end{equation}

It would be enough to show that there is at least one configuration of our general solution whose Casimir invariant cannot be matched by the axial-torsion one (\ref{CGT invariant}). Choosing the particular configuration in our solutions for which $f(r)$ is the same as in \cite{Canfora:2007xs} and both $\chi_p$ vanish. As a consequence,
\begin{equation}
\eta = \frac{r}{f}\,\left( \beta-\frac{r^2}{\ell^2}\right)  \,.
\end{equation}
The constants
$\mu,\beta,C^2$ are arbitrary, so that $\psi_r$ and $\psi_t$ do not vanish in general (unless $\mu=\beta=C^2$), and we get
\begin{align}
\chi_r &  =0\,,\qquad \chi_t =0\,, \qquad\psi_r =\frac{r}{f^2}\, (\beta-\mu)  \,,\nonumber\\
\psi_t &  = r \sqrt{\left(\frac{r^2}{\ell^2} - \beta\right)^2 - \left(\frac{r^2}{\ell^2} - C^2 \right) \left(\frac{r^2}{\ell^2}- \mu\right)}\,.
\end{align}
We observe that the Casimir invariant is clearly different from that in Eq.(\ref{CGT invariant}), as it reads
\begin{equation}
F_{\text{AdS}}^2 =\frac{12}{r^4}\frac{\left(4C^6 + \mu^2C^2 - 2\beta\mu C^2 - \beta^2C^2- 8\beta C^4 + 5\mu C^4 + 2\beta^3 - \mu \beta^2\right) \frac{r^2}{\ell^2} - 4C^4(C^2\mu-\beta^2)}{\left(2\beta - C^2 - \mu \right) \frac{r^2}{\ell^2}+ C^2\mu-\beta^2} \,.
\end{equation}
When $\mu=\beta$, the factors $\mu-C^2$ cancel out and we get
\begin{equation}
F_\text{AdS}^2= \frac{12}{r^4}\frac{\left(\mu^2 - 4C^4 - \mu C^2 \right) \frac{r^2}{\ell^2}+ 4C^4 \mu} {\frac{r^2}{\ell^2}-\mu}\,,
\end{equation}
and in the limit $\mu=C^2$ we obtain
\begin{equation}
F_{\text{AdS}}^{2}=-\frac{48C^{4}}{r^{4}}\,.
\end{equation}
Thus, the invariant takes the same value as in (\ref{CGT invariant}) only for $\mu=C^{2}$. In contrast, when the $\psi_p$  components are non-vanishing, the two Casimir invariants clearly have different forms, showing that the configuration of \cite{Canfora:2007xs} and the one discussed here are physically inequivalent. Switching off the $\psi$s transforms one solution smoothly into the other.

%%%%%%%%%%%%%%%%%%%%%%%%%%%%%%%%
\section{Local symmetries} \label{Local symm}          %   5    %
%%%%%%%%%%%%%%%%%%%%%%%%%%%%%%%%

It is natural to expect that the presence of three arbitrary functions in the general solution (\ref{General1}) and (\ref{General2}) are the consequence of a gauge symmetry. This symmetry cannot be a restriction of the gauge transformation $\mathbf{A}' = g^{-1} (\mathbf{A} + d)g$ that preserves the form of the spherically symmetric ansatz $\mathbf{A}$. In Appendix \ref{BH ansatz}, it is shown that the infinitesimal gauge transformations that preserve this ansatz are necessarily rigid ($g=Const$). Thus, residual gauge symmetries of this kind cannot explain the existence of arbitrary functions in the general solution.

On the other hand, the dynamical structure of CS theories is complex. Namely, these theories are by construction invariant under spacetime
diffeomorphisms and gauge transformations, but one diffeomorphism is always dependent from the gauge transformations in \emph{generic} CS theories, that is, the ones that possess \emph{minimal} number of local symmetries \cite{Banados-Garay-Henneaux}. It may happen, however, that the CS theory is not generic, but it possesses \emph{accidental local symmetries}, where ``accidental'' means that they appear only around some backgrounds.

Because of these special features of the dynamics of CS theories, we suspect that, in our background, there are additional local transformations (different from $\mathbf{\Lambda }$ and $\xi $). The proof is given in the next section using Hamiltonian analysis. Here we take a shortcut by noticing that the functions $\theta $ are arbitrary as long as $f\psi_t\chi_t\eta \neq 0$, so the general solution is insensitive to the infinitesimal changes
\begin{align}
\delta\theta &  =2\sigma(r)\,,\nonumber\\
\delta\theta_{t} &  =2\int dr\,\tau(r)\,,\nonumber\\
\delta\theta_{r} &  =-2\int dr\,\rho(r)+2\int dr\,\int\limits_{0}^{r}
ds\,\tau(s)+2\sigma(r)\,.
\end{align}
This induces the following local transformations of the metric, the electromagnetic field and the torsion components,
\begin{align}
\delta f &  =\frac{\sigma}{f}\,,\nonumber\\
\delta A_t &  =-\frac{k}{C\ell \alpha}\left[ r\sigma^\prime+\frac{2\eta}{rf}\,\sigma
+\frac{rf^2}{\chi_t}\,\tau-\left( r+\frac{f\eta}{r\chi_t}\right)\rho\right]  \,, \nonumber\\
\delta\psi_{r} &  =\frac{r^{2}}{\chi_{t}}\,\tau-\frac{\eta}{f\chi_{t}}\,\rho\,, \label{onshell tr} \\
\delta\psi_{t} &  =\left(  \frac{\psi_{t}}{f^{2}}+\frac{\eta^{2}}{\psi_{t}}\right)\sigma
+\frac{f^{2}\eta}{\psi_{t}\chi_{t}}\,\left(  \rule{0pt}{15pt}r^{2}f\,\tau-\eta\rho\right)  \,,\nonumber\\
\delta\chi_{r} &  =\frac{r^{2}}{\psi_{t}}\,\sigma^{\prime\prime}
-\chi_{r}\left(  \frac{1}{f^{2}}+\frac{\eta^{2}}{\psi_{t}^{2}}\right) \sigma
+\frac{r^{2}}{\psi_{t}}\left(  1-\frac{f^{3}\eta\chi_{r}}{\psi_{t}\chi_{t}}\right)\tau
+\frac{f^{2}\eta^{2}\chi_{r}}{\psi_{t}^{2}\chi_{t}}\,\rho
-\frac{r^{2}}{\psi_{t}}\,\rho^{\prime}\,,\nonumber\\
\delta\chi_{t} &  =\rho\,, \nonumber
\end{align}
where we used the auxiliary expression
\begin{equation}
\delta\eta=\frac{\eta}{f^2}\,\sigma + \frac{r^2 f}{\chi_t}\,\tau - \frac{\eta}{\chi_t}\,\rho\,.
\end{equation}
Direct calculation shows that these transformations, with local parameters $\sigma(r)$, $\tau(r)$ and $\rho(r)$, leave the field equations invariant,
\begin{align}
\delta\mathcal{T}(r) &  =0\,,\nonumber\\
\delta\mathcal{E}(r) &  =\mathcal{E}(r)\,\frac{\sigma}{f^{2}}\,,\nonumber\\
\delta\mathcal{S}(r) &  =\mathcal{S}(r)\,\frac{\sigma}{f^{2}}\,.
\end{align}
Also, the transformations are Abelian because $[\delta_1,\delta_2] =0$ upon acting on any field, so the operators that generate them must also commute. This new unexpected on-shell symmetry $U(1)\times U(1)\times U(1)$ cannot be a Cartan subgroup of $SO(2,4)\times U(1)$ because we already showed that there are no residual gauge symmetries.

In the next section we will prove that the Hamiltonian is (off-shell) invariant under 4-parameter local symmetry that on-shell reduces to the 3-parameter transformations Eq.(\ref{onshell tr}).

%%%%%%%%%%%%%%%%%%%%%%%%%%%%%%%%%%%%%%%
\section{Hamiltonian analysis}   \label{Hamiltonian analysis}            %  6   %
%%%%%%%%%%%%%%%%%%%%%%%%%%%%%%%%%%%%%%%

We shall work in a radial minisuperspace reduction of CS AdS gravity, in which the $r$ coordinate plays the role of time. This is a consistent truncation of the theory involving only relevant degrees of freedom. In practice, it means plugging in an ansatz of the fields directly in the action and studying its effective behavior. The first order CS action is expected to remain linear in velocities also in the approximation.

The validity of the approximation is guaranteed by the theorems of Palais \cite{Palais:1979rca}. It can be successfully applied to a gravitation theory possessing highly symmetric solutions \cite{Deser:2003up} provided the components $g_{tt}$ and $g_{rr}$ are kept independent since, as noted in \cite{Deser:2004yh}, assuming $g_{tt}g_{rr}=-1$ can lead to inconsistencies. In what follows we will check explicitly that our effective action gives rise to the same equations of motion as the original one.

%%%%%%%%%%%%%%%%%%%%%%%%%%%%%%%%%%%%%%%%%%%%%%%%%%%%%%%%%%%%%%%%%%%%%%%%%%%%%
\subsection{Effective action and equations of motion}
%%%%%%%%%%%%%%%%%%%%%%%%%%%%%%%%%%%%%%%%%%%%%%%%%%%%%%%%%%%%%%%%%%%%%%%%%%%%%

We generalize the metric ansatz (\ref{metric_ansatz}) in the coordinates
$x^\mu=(t,r,x^m)$, $m=2,3,4$, so that $g_{tt}=-h^2f^2$ and $g_{rr}=1/f^2$ describe independent metric fields of
a static, spherically symmetric, planar black hole,
\begin{equation}
ds^2=g_{\mu\nu}\,dx^\mu dx^\nu=-h^2(r)f^2(r)\,dt^2+\frac{dr^2}{f^2(r)}+r^2\delta_{mn}\,dx^m dx^n\,.
\end{equation}
The vielbein is given by
\begin{equation}
e^{0}=hf\,dt\,,\qquad e^{1}=\frac{dr}{f}\,,\qquad e^{i}=r\,\delta_{m}^{i}\,dx^{m}\,,\label{e}
\end{equation}
and the spin connection reads
\begin{equation}
\begin{array}
[c]{llll}
\omega^{01}= & \omega\,dt-\chi dr\ ,\qquad & \omega^{1i}= & \nu\,dx^i\,, \medskip\\
\omega^{0i}= & -\psi\,dx^i\,, & \omega^{ij}= & -\varphi\,\epsilon_{\ \ k}^{ij}\,dx^k\,.
\end{array}
\label{w}
\end{equation}
The components of $\omega^{ab}$ are fundamental fields in the first order formalism, and they are defined by
\begin{eqnarray}
\omega  &=&f(fh)'-\frac{\chi_t}{h}\,,\qquad \chi=\frac{\chi_r}{h}\,,\qquad \varphi =\frac{\phi }{2r^2}\,,\notag\\
\nu &=&\frac{f}{r}\,(\psi_r-r)\,,\qquad \psi=\frac{\psi_t}{rhf}\,.
\end{eqnarray}
Thus, in this section, the dynamical fields  are $\{ \varphi, \psi, \nu, \omega, \chi\}$
instead of the torsion components $\{ \phi, \psi_t, \psi_r, \chi_t, \chi_r \}$,
and their dimensions in the length units are $1/L$ for $\omega$ and dimensionless for all other fields.

Imposing the spherically symmetric ansatz on the electromagnetic field
\begin{equation}
A=A_t(r)\,dt\,, \label{ansatz A}
\end{equation}
we find that the electromagnetic kinetic term vanishes, and the interaction $L_{\text{int}}=\alpha\,dB\wedge A$ can be calculated using the identity showed in Appendix \ref{BH ansatz},
\begin{equation}
\frac{1}{2}\,R^{ab}R_{ab}+\frac{1}{\ell^2}\,\left( R^{ab}e_a e_b-T^a T_a\right)  =dB\,,
\end{equation}
where
\begin{equation}
B=\varphi\left(\frac{1}{3}\,\varphi^2+\nu^2-\psi^2
-\frac{r^2}{\ell^2}\right)  \epsilon_{knm}\,dx^k\wedge dx^n\wedge dx^m\,.
\end{equation}
Plugging in the ansatz (\ref{e})-(\ref{ansatz A}) in the CS action, we obtain the effective action
\begin{align}
\mathcal{I}_\text{eff} &  =\frac{6k}{\ell}\,\int dr\,\left[
\left(-\omega'r+\frac{\omega\nu}{f}+hf\chi\psi+hf\,\nu'+\frac{hr}{\ell^{2}}\right)
\left( \psi^{2}-\varphi^{2}-\nu^{2}\right)  \right.  \nonumber\\
&  +\omega r\left(  \nu^{2}-\psi^{2}\right)^{\prime}-2hf\nu\,\varphi\varphi'
-\frac{\omega^{\prime}r^{3}}{3\ell^{2}}+\frac{\omega\nu r^{2}}{f\ell^{2}}
+\frac{hfr^{2}}{\ell^{2}}\,( \chi\psi+\nu') \nonumber\\
&  \left.  +\frac{hr^{3}}{\ell^{4}}-\frac{\alpha \ell}{k}\,\left(
\frac{1}{3}\,\varphi^{2}+\nu^{2}-\psi^{2}
-\frac{r^{2}}{\ell^{2}}\right) \varphi A_{t}^{\prime}\right]  \,,\label{for_variations}
\end{align}
where $\mathcal{I}_\text{eff}=I_\text{eff}/\text{Vol}(\partial {\cal M})$ is the action per unit time and unit volume of transversal section.
It can be further simplified, up to a boundary term, as
\begin{align}
\mathcal{I}_\text{eff} &  =\frac{6k}{\ell}\int dr\,\left[ \left( \frac{\omega\nu}{f}
+hf\chi\psi+hf\nu^\prime+\frac{hr}{\ell^2}+\omega\right)
\left( \psi^2-\varphi^2-\nu^2+\frac{r^2}{\ell^2}\right) \right.  \nonumber\\
&  -\left.  2\left(  r\omega+hf\nu\right)  \,\varphi\varphi^{\prime}
-\frac{\alpha \ell}{k}\,\left(  \frac{1}{3}\,\varphi^{2}+\nu^{2}
-\psi^{2}-\frac{r^{2}}{\ell^{2}}\right) \varphi A_{t}^{\prime}\right]  \,.\label{Ieff}
\end{align}

This action leads to the same field equations as non-truncated CS AdS gravity evaluated in the ansatz.
To show this, let us denote
\begin{align}
\mathcal{T}_{1}(r) &  =\psi^{2}-\varphi^{2}
-\nu^{2}+\frac{r^{2}}{\ell^{2}}\,,\nonumber\\
\mathcal{S}_{1}(r,h) &  =\nu\left(  hf\right)  ^{\prime}
-hf\chi\psi-\frac{\nu\omega}{f}+r\omega^{\prime}
-\frac{hr}{\ell^{2}}\,,\nonumber\\
\mathcal{E}_{1}(r,h) &  =\frac{\omega\nu}{f}+hf\chi
\psi+hf\nu^{\prime}+\frac{hr}{\ell^{2}}+\omega
+\frac{\alpha \ell}{k}\,\varphi A_{t}^{\prime}\,.\label{eom_h}
\end{align}
Then the equations of motion that render the effective action (\ref{Ieff}) stationary are
\begin{align}
\delta h &  :\quad 0=\left( \chi\psi+\nu^{\prime}
+\frac{r}{\ell^{2}f}\right)  \mathcal{T}_{1}(r)-2\nu\,\varphi \varphi^{\prime}\,,\nonumber\\
\delta f &  : \quad 0=\left(\chi\psi+\nu^{\prime}-\frac{\omega\nu}{hf^{2}}\right)  \mathcal{T}_{1}(r)
-2\nu\,\varphi\varphi^{\prime}\,,\nonumber\\
\delta A_{t} &  :\quad 0=\left(\mathcal{T}_{1}(r)\,\varphi+\frac{2}{3}\,\varphi^{3}\right)'\,,
\end{align}
and for the torsion components
\begin{align}
\delta\chi &  :\quad 0=hf\,\psi\mathcal{T}_{1}(r)\,,\nonumber\\
\delta\varphi &  : \quad 0=\mathcal{S}_{1}(r,h)\,\varphi
+\frac{\alpha \ell}{2k}\,\mathcal{T}_{1}(r)A_{t}^{\prime}\,,\nonumber\\
\delta\psi &  :\quad 0=hf\,\chi\mathcal{T}_{1}(r)+2\mathcal{E}_{1}(r,h)\,\psi\,,\nonumber\\
\delta\nu &  : \quad 0=-\frac{\omega}{2f}\,\mathcal{T}_{1}(r)+\frac{1}{2}\left(  hf\,\mathcal{T}_{1}(r)\right)'
+hf\,\varphi\varphi'+\mathcal{E}_{1}(r,h)\,\nu\,,\nonumber\\
\delta\omega  &  :\quad 0=\left(  1+\frac{\nu}{f}\right)
\mathcal{T}_{1}(r)-2r\varphi\varphi'\,.
\end{align}

In the particular case $h=1$, we get
\begin{align}
h &  =1\,,\qquad \varphi^\prime=0 \; \Rightarrow \; \varphi=C \,,\nonumber\\
0 &  =\mathcal{T}_1(r)=\psi^{2}-\varphi^{2}
-\nu^{2}+\frac{r^{2}}{\ell^{2}}\,,\nonumber\\
0 &  =\mathcal{E}_{1}(r,1)=f{\chi}{\psi}
+f\nu'+\frac{r}{\ell^{2}}+\omega+\frac{\omega\nu}{f}
+\frac{\ell \alpha}{k}\,\varphi A_{t}^{\prime}\,,\nonumber\\
0 &  =\mathcal{S}_{1}(r,1)=\nu f^{\prime}-f{\chi}{\psi}-\frac{\nu\omega}{f}+r\omega'-\frac{r}{\ell^{2}}\,,
\end{align}
concluding that the above system indeed reproduces the CS field equations (\ref{C}), $\mathcal{T}=-r^2\mathcal{T}_1$, $\mathcal{E}=-r^2f\mathcal{E}_1$ and $\mathcal{S}=rf\mathcal{S}_1$ (see Eqs.(\ref{Tau}), (\ref{E(r)}) and (\ref{la320})).

%%%%%%%%%%%%%%%%%%%%%%%%%%%%%%%%%%%%%%
\subsection{Constraint structure \label{Constraints}}     %   6 .   2  %
%%%%%%%%%%%%%%%%%%%%%%%%%%%%%%%%%%%%%%
As mentioned before, keeping the metric functions $h(r)$ and $f(r)$ independent ensures the  validity of the minisuperspace approximation, as they usually describe dynamically propagating degrees of freedom. In the considered CS gravity, however, the metric component $h(r)$ is not dynamical, that is, the field equations do not imply $h=1$. As shown in  Appendix \ref{with h}, $h$ can change arbitrarily due to a one-parameter local transformation. Thus, $h=1$ can be chosen as a gauge fixing. The effective action also shows that setting $h=1$ gives another consistent truncation of the action in the sense that it has an extremum on the correct equations of motion. From now on, therefore, we will set $h=1$, but in Appendix \ref{with h} we prove that the results are the same as for general $h(r)$.

The generalized coordinates $q_s(r)$ and their corresponding conjugate momenta $p^s(r)=\delta\mathcal{I}_\text{eff}/\delta q'_s$ define 14-dimensional phase space $\Gamma$,
\begin{equation}
q_s=\{ f,A_t,\varphi,\psi,\nu,\omega,\chi\}\,,\qquad p^s=\{ p_f,p_A,p_\varphi,p_\psi,p_\nu,p_\omega,p_\chi\}\,.
\end{equation}
Their fundamental Poisson brackets (PB) taken at the same radial distance $r$,
\begin{equation}
[ q_s,p^{s'} ]=\delta_s^{s'}\,.
\end{equation}
Since the action (\ref{Ieff}) is first order (it does not contain second derivatives), all momenta become algebraic functions of the coordinates, giving rise to the primary constraints
\begin{equation}
\begin{array}
[c]{llll}
C_f & =p_f \approx0\,, & C_\nu & = p_\nu- \frac{6k}{\ell}f\,\mathcal{T}_1 \approx0\,,\medskip\\
C_\psi & =p_\psi \approx0\,, & C_\varphi & = p_\varphi + \frac{12k}{\ell}\, (r\omega + f\nu ) \,\varphi \approx0\,,\medskip\\
C_\chi & =p_\chi \approx0\,,\medskip & C_A & = p_A - 4\alpha \,\varphi ^3 - 6\alpha \,\varphi \mathcal{T}_1 \approx0\,.\\
C_{\omega} & =p_{\omega}\approx0\,,\medskip\qquad &  &
\end{array}
\end{equation}
The constraints $C_s(q,p,r)\approx 0$ define the primary constraint surface $\Sigma_\text{P}$.

Let us recall that the weak vanishing of some smooth, differentiable function $X(q(r),p(r),r)$ means that it vanishes on the constraint surface, that is, $X\approx 0 \Leftrightarrow X|_{\Sigma _\text{P}}=0$. In order the equality to become strong, one needs both $X$ and $X^{\prime}$ to vanish on the constraints surface, i.e.,  $X = 0 \Leftrightarrow X,X'|_{\Sigma _\text{P}}=0$. A strong and weak equalities are equivalent up to a linear combination of the constraints, that is, $X\approx 0 \Leftrightarrow X=u^s C_s$ .

The canonical Hamiltonian obtained from the effective action (\ref{Ieff}) has the form
\begin{equation}
\mathcal{H}_\text{C}(p,q,r)= p^s q'_s-\mathcal{L}_\text{eff}  \approx - \frac{6k}{\ell}\,\left(  \frac{\omega\nu }%
{f}+f\chi \psi +\frac{r}{\ell^{2}}+\omega\right)  \mathcal{T}_1\,,
\end{equation}
and it naturally leads to the definition of the total Hamiltonian that also depends on Lagrange multipliers $u^s(r)$,
\begin{equation}
\mathcal{H}_\text{T}(p,q,u,r)=\mathcal{H}_\text{C}+u^{s}C_{s}\,.
\end{equation}

Consistency requires that all constraints remain vanishing throughout their evolution,
\begin{equation}
 C'_q = \frac{\partial C_q}{\partial r}+[ C_q,\mathcal{H}_\text{T}] \approx 0\,.
\end{equation}
These conditions give rise either to secondary constraints, or they determine some multipliers $u^s$. Choosing the branch with $f\phi\psi\chi\neq0$, we find that $C'_\chi \approx 0$ leads to a secondary constraint
\begin{equation}
\mathcal{T}_{1}\approx0\,,\label{Tau constraint}
\end{equation}
whereas $C_{\omega}^{\prime}\approx0$ solves a multiplier
\begin{equation}
u^\varphi =0\,.
\end{equation}
The constraint $C_f$ does not change along $r$, and $C'_\psi \approx0$, $C'_A \approx0$, $C'_\nu \approx0$ and $C'_\varphi \approx0$ determine three Hamiltonian multipliers,
\begin{align}
u^A & =-\frac{k}{\alpha \ell}\,\varphi ^{-1}\left(\frac{\omega\nu }{f} + f \chi \psi  + \frac{r}{\ell^2} + \omega + f\, u^\nu \right)\,,\nonumber\\
u^\psi & = \psi ^{-1} \left(\nu u^\nu - \frac{r}{\ell^2}\right)  \,,\nonumber\\
u^\omega & =\frac{1}{r} \left(\frac{\omega\nu }{f} +f \chi \psi  + \frac{r}{\ell^2} - \nu \,u^f \right)  \,.
\end{align}
Finally, the secondary constraint $\mathcal{T}_1$ does not change along $r$. We conclude that the final constraint surface, $\Sigma$, is defined by the sets
\begin{align*}
\text{Primary constraints}    &: \qquad \{  C_f, C_\psi, C_\chi, C_\omega, C_\nu, C_\varphi  , C_A \}\,,\\
\text{Secondary constraints}  &: \qquad \{ \mathcal{T}_1 \}\,.
\end{align*}

In order to identify the local symmetries, we have to separate first class constraints. By definition, first class constraints $G_a \approx 0$ commute with all other constraints on the surface $\Sigma$, while second class constraints $S_\alpha\approx 0$ have nonsingular PBs on $\Sigma$.

A separation between first and second class constraints $(G_a, S_\alpha)$ has to be achieved by redefinition of constraints so that the surface $\Sigma$ remains unchanged. Hence, the first class constraints $G_a$ are obtained as
\begin{align}
G_f &  = f \left( C_{f}-\frac{\nu }{r}\,C_{\omega} \right)\,,\nonumber\\
G_\nu &  =C_\nu +\frac{\nu }{\psi }\,C_{\psi}-\frac{k}{\alpha \ell}\,\frac{f}{\varphi }\,C_{A}\,,\nonumber\\
G_\tau &  =f \left( -\frac{6k}{\ell}\,\mathcal{T}_{1}-\frac{k}{\alpha \ell}\frac{1}{\varphi }\,C_{A}
+\frac{1}{r}\,C_{\omega}\right) \,,\nonumber\\
G_\chi &  =C_{\chi}\,.
\end{align}
They satisfy the first class subalgebra
\begin{equation}
[G_\nu,G_f]  =G_\tau\,,\qquad [ G_\tau, G_f ] =G_\tau\,.  \label{FC commutators}
\end{equation}

The second class constraints $S_\alpha$ have the form
\begin{align}
S_\varphi & =\varphi C_\varphi  \,,\qquad  S_\psi  =\frac{1}{\psi}\,C_\psi\,, \nonumber\\
S_\omega & =\frac{f}{r}\,C_\omega\,,\qquad S_A  = C_A\,,
\end{align}
and their PBs define the symplectic matrix $\Omega_{\alpha\beta}=\{ S_\alpha, S_\beta\}$ that is invertible on $\Sigma$,
\begin{align}
[ S_\varphi,S_\omega]   &  =\frac{12k}{\ell}\,f\varphi^2\,,\nonumber\\
[ S_\varphi,S_A]   &  =6\alpha \,\varphi\mathcal{T}_1\,,\nonumber\\
[ S_\psi,S_A]   &  =12\alpha \,\varphi\,.
\label{SC commutators}
\end{align}
It can be seen that $\Omega_{\alpha\beta}$ is indeed non-singular, $\det\Omega|_\Sigma= 144\, \alpha k f\, \varphi^3/\ell \neq 0 $.

The first and second class constraints are easily distinguished if they commute with each other on  $\Sigma$, and this is in fact the case,
\begin{align}
[  S_{\varphi},G_{\nu}]   &  =G_{\tau}-S_{\omega}\,,\nonumber\\
[  S_{\varphi},G_{\tau}]   &  =G_{\tau}-S_{\omega}\,,\nonumber\\
[  S_{\omega},G_{f}]   &  =S_{\omega}\,.
\label{FC-SC commutators}
\end{align}

%At the end of this section, let us show that Hamilton's equations of motion are equivalent to the
%Euler-Lagrange equations of motion. Using $q'_s=[q_s,\mathcal{H}_\text{T}] $, we obtain equations of motion
%for the fields
%\begin{equation}
%\begin{array}
%[b]{llll}
%f^{\prime} & =u^{f}\,, & \psi ^{\prime} & =u^{\psi}
%=\psi ^{-1}\left(  \nu \,u^\nu -\frac{r}{\ell^{2}}\right)  \,,\medskip\\
%\chi ^{\prime} & =u^{\chi}\,, & \omega^{\prime} & =u^{\omega}=\frac
%{1}{r}\left(  \frac{\omega\nu }{f}+f\chi \psi
%+\frac{r}{\ell^{2}}-\nu u^{f}\right)  \,,\medskip\\
%\nu ^{\prime} & =u^\nu \,, & A_{t}^{\prime} & =u^{A}=\frac
%{k}{3a\ell}\,\varphi ^{-1}\left(  \frac{\omega\nu }{f}
%+f\chi \psi +\frac{r}{\ell^{2}}+\omega+f\,u^\nu \right)
%\,.\medskip\\
%\varphi ^{\prime} & =u^\varphi =0\,,\qquad &  &
%\end{array}
%\label{Hamilton EQ}
%\end{equation}

%The Euler-Lagrange equation $\mathcal{T}_1(r)=0$ corresponds to the
%secondary Hamiltonian constraint. Other two Euler-Lagrange equations are
%obtained by combining the Hamilton's equations for $f'$, $\omega'$, $\nu'$
%and $A'_t$. Indeed, by a direct replacement, we find $\mathcal{S}_1(r)=0$
%and $\mathcal{E}_1(r)=0$.

Adding the secondary constraint $G_\tau$ with the multiplier $U^\tau$ to the total Hamiltonian, plugging in all solved multipliers $u^s$ and redefining unsolved multipliers as $U^f=u^f/f\,,U^\nu=u^\nu $ and $U^\chi=u^\chi$, the extended Hamiltoniani is obtained
\begin{equation}
\mathcal{H}_{\text{E}}=\mathcal{H}_{0}+U^{a}G_{a}\,,
\end{equation}
where from now on $U^a(r)$ are field-independent Lagrange multipliers. The new canonical Hamiltonian reads
\begin{eqnarray}
\mathcal{H}_0 &= &-\frac{6k}{\ell}\,\left(\frac{\omega \nu }{f}
+f\chi \psi + \frac{r}{\ell^2}+ \omega \right) \mathcal{T}_1 +\left(\frac{\omega \nu}{f}+ f\chi \psi + \frac{r}{\ell^2}\right) \,\frac{C_\omega}{r}  \nonumber \\
&&-\frac{r}{\ell^2\psi }\,C_\psi - \frac{k}{\alpha \ell}\,\left(\frac{\omega \nu }{f} + f\chi \psi  + \frac{r}{\ell ^2} +\omega \right) \frac{C_A}{\varphi }\,.
\end{eqnarray}
%Notice that $\mathcal{H}_0\approx 0$,  that is a typical feature of gravity actions.

Hamilton's equations can be shown to be equivalent to the Euler-Lagrange ones. Using $q'_s\approx [q_s,\mathcal{H}_0]+U^a[f,G_a]$,
\begin{equation}
\begin{array}[b]{llll}
f' & \approx fU^f\,, & \psi '  & \approx \psi ^{-1}\left(\nu U^\nu - \frac{r}{\ell^2}\right)\,,\medskip  \\
\chi ' & \approx U^\chi \,, & \omega'  & \approx \frac{\omega \nu}{rf}+ \frac{f\chi \psi }{r}
+ \frac{1}{\ell^2} + \frac{f}{r}\,\left(U^\tau -\nu U^f \right)\,,\medskip  \\
\nu ^{\prime } & \approx U^{\nu }\,, & A_{t}^{\prime } &
\approx -\frac{k}{\alpha \ell }\,\varphi ^{-1} \left(\frac{\omega \nu }{f}+f\chi \psi +\frac{r}{\ell ^2}+\omega +f U^\nu + f U^\tau \right) \,.\medskip  \\
\varphi ' & \approx 0\,,\qquad  &  &
\end{array}
\label{Hamilton EQ}
\end{equation}
By direct replacement of the above expressions, all multipliers cancel out and the Euler-Lagrange equations $\mathcal{T}_1(r) = 0$, $\mathcal{S}_1(r)=0$ and $\mathcal{E}_1(r)=0$ are reproduced.

%%%%%%%%%%%%%%%%%%%%%%%%%%%%%%
\subsection{Counting of degrees of freedom}     %   6  . 3  %
%%%%%%%%%%%%%%%%%%%%%%%%%%%%%%
Dirac's procedure allows counting the physical degrees of freedom in a theory, the ones that remain after gauge fixing of all local symmetries and after elimination of non-physical variables due to second class constraints. In a theory with $n$ generalized coordinates, $n_1$ first class constraints and $n_2$ second class constraints, the number of degrees of freedom is $\mathcal{F}=n-n_1-\frac{1}{2} n_2$.

In our case there are $n=7$ fundamental fields $q_s=\{f,A_t,\varphi,\psi,\nu,\omega,\chi \}$ and $n_1=4$, $n_2=4$ constraints, leading to one degree of freedom,
\begin{equation}
\mathcal{F}=1\,.
\end{equation}

On the other hand, the degrees of freedom can be counted for \emph{generic} CS gauge theories in $D=2k+1$ for a non-Abelian Lie algebra with $N$ generators \cite{Banados-Garay-Henneaux}: the theory has $N$ first class constraints $\mathcal{G}_M\approx 0$ (generators of gauge transformations) and a set of $2kN$ mixed first and second class constraints $\phi_M^{\bar{m}}\approx 0$, where $\bar{m}=(t,m)$ denotes the boundary spacetime indices. In general, there is no simple algorithm to separate first and second class constraints among the $\phi_M^{\bar{m}}$. The symplectic form is
\begin{equation}
\{\phi_M^{\bar{m}}, \phi_N^{\bar{n}}\} =\Omega_{MN}^{\bar{m}\bar{n}} :=
\epsilon^{\bar{m}\bar{n}\bar{m}_1\bar{n}_1\cdots \bar{m}_{k-1}\bar{n}_{k-1}} g_{MNK_1\cdots K_{k-1}}
\,F_{\bar{m}_1\bar{n}_1}^{K_1}\cdots F_{\bar{m}_{k-1}\bar{n}_{k-1}}^{K_{k-1}}\, .
\end{equation}
In general, the number of first class among the $\phi_M^{\bar{m}}$ corresponds to the number of zero modes of the $2kN\times 2kN$ matrix $\Omega_{MN}^{\bar{m}\bar{n}} $, while its rank corresponds to the number of second class constraints. As shown in  Ref.\cite{Banados-Garay-Henneaux}, $\Omega$ has always at least $2k$ zero modes, $\mathcal{H}_{\bar{m}}= F_{\bar{m}\bar{n}}^M\,\phi_M^{\bar{n}}$, which generate diffeomorphisms in the transverse section, while the radial diffeomorphism is not an independent symmetry.

Clearly, the rank of $\Omega$ and the number of its zero modes depend on the values of the components $F_{\bar{m}\bar{n}}^K$ at each point in spacetime. A generic configuration is, by definition, one in which the rank of  $\Omega$ is the maximum possible and therefore the number of local symmetries is minimal. In such case, $\Omega$ has exactly $2k$ zero modes, and the number of degrees of freedom is the maximum a CS theory can have. In those sectors, there are $n_1=N+2k$ first class constraints $(\mathcal{G}_M,\mathcal{H}_{\bar{m}})$ and $n_2=2kN-2k$ second class constrains corresponding to $\phi_M^{\bar{n}}$ where the $\mathcal{H}_{\bar{m}}$ have been eliminated. Applying the Dirac formula for $n=2kN$ gauge fields $A_{\bar{m}}^M$ (without the Lagrange multipliers $A_t^M$), one obtains
\begin{equation}
\mathcal{F}_\text{CS generic}=kN-k-N\,.
\end{equation}
An explicit separation of first and second class constrains in a generic sector of a $G\times U(1)$ CS theory was done in Ref.\cite{Banados-Garay-Henneaux}, however, the separation for other CS theories is not known in general.

In our five-dimensional case ($k=2$), the Lie group AdS$_5\times U(1)$ has $N=16$ generators, so the generic CS AdS gravity has $\mathcal{F}_\text{CS generic}=14$ degrees of freedom, that is much more than what we proved to exist in the background of Section 3, $\mathcal{F}=1$. We conclude that the symmetric background, whose symplectic 2-form $\Omega_{MN}=g_{MNK}\,F^K$ has components
\begin{align}
\Omega_{11} & =\beta \,F\,,\qquad \Omega_{1a}=-\frac{\alpha}{\ell}\,T_a\,,
\qquad \Omega_{1[ab]}  =\alpha \,F_{ab}\,,\qquad \Omega _{ab}=-\alpha \,\eta _{ab}\,F\,,  \nonumber\\
\Omega_{a[bc]} &=-\frac{k}{2}\,\epsilon _{abcde}\,F^{de}\,, \qquad
\Omega_{[ab][cd]} =\alpha \,(\eta_{ac}\eta _{bd}-\eta_{ad}\eta_{bc})\,F-\frac{k}{\ell }\,\epsilon_{abcde}\,T^e\,,
\end{align}
is not in a generic sector of CS AdS gravity, but it contains additional zero modes, related to the accidental local symmetries discussed in Section \ref{Local symm}. In the next section, we study these symmetries in the context of Hamiltonian formalism.

%%%%%%%%%%%%%%%%%%%%%%%%%%%
\section{Hamiltonian local symmetries}         %  7  %
%%%%%%%%%%%%%%%%%%%%%%%%%%%
Dirac's method provides a systematic way to identify local symmetries of the Hamiltonian system. A symmetry with local parameters $\lambda^a(r)=(\lambda^f,\lambda^\tau,\lambda^\nu,\lambda^\chi)$ is obtained from a generator  $G[\lambda]$ constructed from first class constraints,
\begin{equation}
G[\lambda]=\lambda^a G_a\, .\label{G(lambda)}
\end{equation}
Then, local transformations of the form
\begin{equation}
\delta q_s=[q_s,G[\lambda] ] \,,\qquad \delta p^s=[p^s,G[\lambda] ]\,,
\end{equation}
leave the Hamiltonian $\mathcal{H}_\text{E}$ invariant. Explicitly, the fundamental fields change as
\begin{equation}
\begin{array}
[b]{lll}
\delta f & =f\lambda^f\,,\medskip & \delta\psi=\nu\,\psi^{-1}\lambda^\nu\,,\\
\delta\chi & =\lambda^\chi\,,\medskip & \delta\omega=\frac{f}{r}\left(\lambda^\tau -\nu\,\lambda^f\right),\\
\delta\nu & =\lambda^\nu\,,\medskip\qquad & \delta A_t =-\frac{k}{\alpha \ell}\,f\varphi^{-1}\left(\lambda^\tau+\lambda^\nu\right).\\
\delta\varphi  & =0\ , &
\end{array}  \label{Hamiltonian local}
\end{equation}
This four-parameter local symmetry is non-Abelian. On the other hand, the on-shell local symmetry of the Lagrangian presented in Section
\ref{Local symm} is three-parameter one and Abelian. A  relation between the Hamiltonian and Lagrangian symmetries is given by Castellani's procedure \cite{Castellani}, where a difference occurs when there are secondary constraints that are a part of the symmetry generator. In fact, for each secondary first class constraint, the Lagrangian generator involves one first derivative of local parameters associated to primary first class constraints. These derivatives of Lagrangian parameters are treated as \emph{independent} local parameters in the Hamiltonian procedure, which means that Hamiltonian symmetries always possess larger number of local parameters when secondary first class constraints exist.

Similar situation happens in Maxwell electrodynamics, where the first class constraints generate the Hamiltonian local transformations $\delta A_t =\varepsilon$ and $\delta A_i =\partial_i\lambda$ with two independent parameters $\varepsilon$ and $\lambda$, whereas the Lagrangian transformation law, $\delta A_\mu =\partial_\mu\lambda$, relates these parameters as $\varepsilon=\dot{\lambda}$.

In our case, there is one secondary constraint, $\mathcal{T}_1\approx 0$ and therefore one parameter among the $\lambda^a$s is expected to be a first derivative of the others; as shown below $\lambda^\tau$ is that parameter. We are interested in showing the on-shell equivalence between the Lagrangian transformations (\ref{onshell tr}) and the Hamiltonian transformations (\ref{Hamiltonian local}). Thus, it is not necessary to apply Castellani's method in full, it is enough to check invariance of the Hamiltonian equations. To this end, we first write the transformations (\ref{onshell tr}) in terms of more familiar variables $( f,\psi ,\nu ,\omega,\chi ,\varphi ,A_{t}) $. Additionally, we change the local parameters as $(\sigma,\rho,\tau) \rightarrow(\sigma,\gamma,\xi)$, where $\gamma=\sigma' - \rho$ and $\xi= (rf\,\tau-\nu \,\rho)/(ff'- \omega)$. Then the Lagrangian local transformations become
\begin{align}
\delta_L f &  =\frac{\sigma}{f}\,,\qquad\qquad\delta_L\omega=\gamma\,,\qquad\qquad \delta_L\varphi =0\,,\nonumber\\
\delta_L\psi  &  =\frac{\nu }{\psi }\, \delta_L\nu \,,\qquad \delta_L\nu =\frac{\nu }{f^2} \sigma + \xi\,, \nonumber\\
\delta_L\chi  &  =-\frac{\chi }{f^2}\left(1+\frac{\nu ^2 }{\psi ^2}\right) \sigma + \frac{\nu \left(\sigma' - \gamma\right)}{f^2 \psi } + \frac{r}{f\psi }\,\gamma' +\frac{\psi \left(ff' - \omega\right) -f^2 \nu \chi }{f^2 \psi ^2}\,\xi\,,\nonumber\\
\delta_L A_t &  =-\frac{k}{C\ell \alpha}\left(r\gamma + \frac{2\nu }{f}\,\sigma+f\xi\,\right)  \,. \label{Lag symm}
\end{align}
Although originally they seemed to depend on second derivatives of the parameters, it is explicit from (\ref{Lag symm}) that this dependence is on first derivatives only.

Now we turn to the Hamiltonian local transformations. The extended Hamiltonian equations (\ref{Hamilton EQ}) are invariant
when the multipliers transform as
\begin{equation}
\delta U^f  =(\lambda^f )'\,,\qquad \delta U^\chi  =(\lambda^\chi)'\,, \qquad \delta U^\nu = (\lambda^\nu)'\,,
\end{equation}
and
\begin{eqnarray}
\delta U^{\tau } &=&\left( \lambda ^{\tau }\right) ^{\prime }-\psi \lambda ^{\chi }+\left( U^{f}-\frac{\nu }{rf}-\frac{1}{r}\right)
\lambda ^{\tau }+\left( U^{f}-\frac{\omega }{f^{2}}-\frac{\chi \nu }{\psi }\right) \lambda ^{\nu }  \nonumber \\
&&+\left( \frac{\nu }{r}-\chi \psi +\frac{\nu ^{2}}{rf}+\frac{\omega \nu }{f^{2}}-U^{\nu}-U^{\tau }\right)
\,\lambda ^{f}\,.
\end{eqnarray}

Castellani's method is based on the total Hamiltonian that does not include
secondary constraints, thus a relation between two (physically equivalent) descriptions
in terms of either $\mathcal{H}_\text{T}$ or $\mathcal{H}_\text{E}$
is by setting $U^\tau=0$ and, consistently, $\delta U^\tau=0$. The last condition means that
 $(\lambda^\tau)'$ becomes a linear combination of other parameters,
\begin{eqnarray}
(\lambda^\tau)' &=&\psi \lambda ^{\chi
}+\left( \frac{\nu }{rf}+\frac{1}{r}-U^{f}\right) \lambda ^{\tau
}+\left( \frac{\omega }{f^{2}}+\frac{\chi \nu }{\psi }%
-U^{f}\right) \lambda ^{\nu}  \nonumber \\
&&-\left( \frac{\nu }{r}-\chi \psi +\frac{\nu %
^{2}}{rf}+\frac{\omega \nu }{f^{2}}-U^{\nu}\right) \,\lambda
^{f}\,. \label{Castellani}
\end{eqnarray}

As the last step, we redefine the Hamiltonian local parameters $(\lambda^f,\lambda^\nu,\lambda^\tau) \rightarrow (\sigma ,\gamma ,\xi )  $ as
\begin{align}
\lambda^f & =\frac{\sigma }{f^2}\,,  \nonumber \\
\lambda^\nu& =\frac{\nu}{f^2}\,\sigma +\xi \,,  \nonumber \\
\lambda^{\tau} &  =\frac{\nu }{f^{2}}\ \sigma+\frac{r}{f}\ \gamma\,.
\end{align}
The parameter $\lambda^{\chi}$ is not independent due to the relation (\ref{Castellani}) that gives
\begin{equation}
\lambda ^\chi =-\frac{\chi }{f^2} \left(1 + \frac{\nu ^2}{\psi ^2}\right) \sigma + \frac{\nu (\sigma' -\gamma)}{f^2 \psi } + \frac{r}{f \psi }\,\gamma' + \frac{\psi (ff'-\omega ) -f^2 \nu \chi }{f^2 \psi ^2}\,\xi \,.
\end{equation}
Comparing the Hamiltonian transformations with the Lagrangian ones (\ref{Lag symm}), we confirm that they are all the same. This proves that the effective action indeed possesses accidental local symmetries in the spherically symmetric, static background with flat transverse section.

%%%%%%%%%%%%%%%%%
\section{Conclusions}     %   7    %
%%%%%%%%%%%%%%%%%

We have presented an Anti-de Sitter (AdS) black hole solutions in five-dimensional Chern-Simons (CS) supergravity. More precisely, we considered charged black holes with flat horizon, which approach locally AdS$_5$ spacetime at large distances. The minimal setup admitting such AdS$_5 \times U(1)$ configurations in the context of CS supergravity was argued to be the theory formulated on the supergroup $SU(2,2|{\mathcal N})$ which, in addition, contains non-Abelian gauge fields and fermionic matter.

We have shown that, in this theory, black hole solutions charged under the $U(1)$ field do exist, provided the spacetime torsion is non-vanishing. Therefore, we analyzed the most general ansatz consistent with the local AdS$_5$ isometries in Riemann-Cartan space. The coupling of torsion in the action resembles that of the universal axion of string theory, and here it appears to be associated to the $U(1)$ field.

We found explicit charged black hole solutions, which may exhibit locally flat horizons as well as horizons with non-vanishing constant curvature. Motivated by the possible relevance for AdS/CFT, we focused our attention on the flat horizon solutions. These geometries appear as torsionfull five-dimensional generalizations of the three-dimensional black hole \cite{BTZ}; although, in contrast to the latter, our five-dimensional black holes do not present constant curvature; in fact, they present a curvature singularity at the origin hidden behind either one or two smooth horizons.

The simplest charged solutions we found exhibit non-vanishing components of the torsion tensor on the horizon three-surface (axial torsion) as well as along off-diagonal directions involving the radial coordinate. These in turn generalize previous ans\"{a}tze studied in the literature, where only axial torsion was considered.

In the generic case, the fall-off behavior turns out to be weaker than the standard Henneaux-Teitelboim asymptotically AdS boundary conditions \cite{HT}. However, despite this weakened asymptotics, the solutions exhibit finite mass and finite Hawking temperature in the generic case. An extremal configuration also exists, for which the two horizons coincide and the Hawking temperature vanishes. In that case the mass also vanishes and the near horizon geometry is AdS$_2 \times \mathbb{R}^3$. There are particular solutions that are conformally flat, reminiscent of the Riegert's solution of conformal gravity \cite{Riegert}.

We also studied more general solutions, allowing for more non-vanishing components of the torsion tensor that do not violate existing isometries of the spacetime. Such solutions, however, exhibit a peculiar feature: they are characterized by arbitrary functions of the radial coordinate that remain undetermined after the field equations are imposed. Such solutions with a ``free geometry'' of spacetime was noticed thirty years ago by Wheeler within Lovelock gravities \cite{Wheeler:1985qd}. This is also a typical feature of CS gravity theories, which are well-known to contain this type of degeneracy in sectors of its phase space.

Having found new asymptotically AdS$_5$ charged black holes with flat horizon, one can't help speculating about possible consequences that such geometries could have in the context of AdS/CFT correspondence. These solutions could, in principle, lead to gravity duals for conformal field theories (CFT) at finite temperature. However, before trying to interpret our results from the holographic point of view, there are two preliminary questions that should be answered. First, is a general question about the role played by torsion in AdS/CFT. This issue has been addressed in the literature, in particular in the context of CS theory in  three \cite{Klemm} and five dimensions \cite{Banados-Miskovic}, where it was argued that torsion induces new sources in a dual CFT, and in the case of higher-order interactions, it can produce a new kind of conformal anomaly \cite{Blagojevic}.

Second, the question is about the propagating degrees of freedom of the theory. Due to the frugality of CS gravity theory in what regards to its local degrees of freedom, we should wonder how many propagating modes the theory actually has about the symmetric sector of solutions we consider. The answer turns out to be quite interesting. In fact, it is the torsion field the one that makes the theory acquire local degrees of freedom, and through a careful analysis of the canonical structure of the theory, we showed that there is only one dynamically propagating mode in the static symmetric sector of its phase space. This result is in contrast with a generic CS AdS gravity with a $U(1)$ field that possesses 14 dynamically propagating modes. Both theories have the same field content, but they are defined around different backgrounds, i.e., in different sectors of phase space. As discussed in \cite{Banados-Garay-Henneaux}, generic theories have maximal number of degrees of freedom (14 in this case), and that means that the missing degrees of freedom are related to an increase in local symmetries.

This last observation is supported by the fact that a general, torsionful, symmetric solution contains a number of indefinite functions of radial coordinate, which produce a three-parameter Abelian on-shell symmetry different from AdS$_5 \times U(1)$. At first sight, an appearance of this additional ``accidental'' symmetry was unexpected. However, its existence is understood through a careful canonical analysis of the effective action stemming from an approximation that keeps only the relevant (symmetric) degrees of freedom switched on. Using this minisuperspace approximation, the Hamiltonian analysis reveals that the symmetric action is indeed invariant under a 4-parameter non-Abelian off-shell symmetry that is not present in the generic phase space region. Comparison with the Lagrangian symmetries confirms that, on-shell, both local transformations match exactly.

The example analyzed here is, therefore, an explicit realization of a non-generic CS gravity. The metric is not a physical field in this sector, even though a particular gauge fixing (i.e., the metric ansatz choice) can make it looks so. Only the knowledge about the existence of accidental symmetries can help to formulate a simple criterion that avoids such unwanted degenerate ans\"{a}tze. As shown here, the simplest way to avoid an unphysical metric is to assume the most general symmetric ansatz and solve it in such a way that there are no indefinite functions associated to it. We used exactly this method to identify two interesting solutions: the one with the axial torsion already known in the literature \cite{Canfora:2007xs}, and a new 2-components torsion solution studied in Section \ref{Charged BH sol}.

%%%%%%%%%%%%%%%%%%
\section*{Acknowledgements}
%%%%%%%%%%%%%%%%%%
The authors thank Max Ba\~{n}ados, Milutin Blagojevi\'c, Branislav Cvetkovi\'c, Natalie Deruelle, Alan Garbarz, Andr\'es Goya, and Julio Oliva for useful discussions. The work of G.G. was supported by grants PIP and UBACyT from CONICET and UBA. This work was also supported by the Chilean FONDECYT Grants No.1110102, No.3130445 and No.1140155.  O.M. is grateful to DII-PUCV for support through the project No.123.711/2011. The Centro de Estudios Cient\'ificos (CECs) is funded by the Chilean Government through the Centers of Excellence Base Financing Program of CONICYT.

\appendix
%%%%%%%%%%%%%%%%%%%%%%%%%%%%%%%%%%%%%%%%%%%%%%%%
\section{Chern-Simons AdS supergravity in five dimensions \label{CS SUGRA}}         % A   %
%%%%%%%%%%%%%%%%%%%%%%%%%%%%%%%%%%%%%%%%%%%%%%%%

The five-dimensional Chern-Simons AdS supergravity is a gauge theory based on a supersymmetric extension of the group $SO(4,2)$, the super unitary group $SU(2,2|{\mathcal N})$ \cite{Chamseddine,Nahm,Strathdee}. Its fundamental field is a gauge connection 1-form
\begin{equation}
\mathbf{A}=A_\mu ^M(x)\,dx^\mu \,\mathbf{G}_M = \frac{1}{\ell }\,e^a \mathbf{J}_a + \frac{1}{2}\,\omega^{ab} \mathbf{J}_{ab} + \mathcal{A}^\Lambda \mathbf{T}_\Lambda  + \left(\bar{\psi}_\alpha ^s \mathbf{Q}_s ^\alpha - \mathbf{\bar{Q}}_\alpha^s \psi_s ^\alpha \right) +A\, \mathbf{T}_{1}\,,
\end{equation}
where $\ell $ denotes the AdS radius. The gauge fields contained in the bosonic sector of theory, that is AdS$_{5}\times SU({\mathcal N})\times U(1)$, are the vielbein ($e^{a}$), the spin connection ($\omega ^{ab}$), the non-Abelian gauge field ($\mathcal{A}^{\Lambda }$) and the Abelian gauge field ($A$). In addition, there are ${\mathcal N}$ gravitini $\psi _{s}$ that are Dirac fields transforming in a vector representation of $SU({\mathcal N})$. When ${\mathcal N}=1$, the non-Abelian generators are absent and the bosonic sector is just AdS$_5\times U(1)$.

The Lie algebra of the bosonic generators is $su(2,2) + su({\mathcal N}) + u(1)$, and the supersymmetry generators extend this algebra as
\begin{equation}
\begin{array}[b]{ll}
\left[ \mathbf{J}_{AB},\mathbf{Q}_{s}^{\alpha }\right] =-\frac{1}{2}
\,\left( \Gamma _{AB}\right) _{\beta }^{\alpha }\,\mathbf{Q}_{s}^{\beta
}\,,\medskip  & \left[ \mathbf{J}_{AB},\mathbf{\bar{Q}}_{\alpha }^{s}\right]
=\frac{1}{2}\,\mathbf{\bar{Q}}_{\beta }^{s}\,\left( \Gamma _{AB}\right)
_{\alpha }^{\beta }\,, \\
\left[ \mathbf{T}_{\Lambda },\mathbf{Q}_{s}^{\alpha }\right] =\left( \tau
_{\Lambda }\right) _{s}^{r}\,\mathbf{Q}_{r}^{\alpha }\,,\medskip  & \left[
\mathbf{T}_{\Lambda },\mathbf{\bar{Q}}_{\alpha }^{s}\right]
=-\mathbf{\bar{Q}}_{\alpha }^{r}\,\left( \tau _{\Lambda }\right) _{r}^{s}\,, \\
\left[ \mathbf{T}_{1},\mathbf{Q}_{s}^{\alpha }\right] =-i\,
\left( \frac{1}{4}-\frac{1}{{\mathcal N}}\right) \,\mathbf{Q}_{s}^{\alpha }\,,\qquad
& \left[ \mathbf{T}_{1},\mathbf{\bar{Q}}_{\alpha }^{s}\right] =
i\,\left( \frac{1}{4}-\frac{1}{{\mathcal N}}\right) \,\mathbf{\bar{Q}}_{\alpha }^{s}\,.
\end{array}
\end{equation}
All generators are anti-Hermitian and the dimension of this superalgebra ${\mathcal N}^{2}+8{\mathcal N}+15$. The AdS indices are denoted by $A=(a,5)$, so that the AdS translations correspond to $\mathbf{J}_{a5}=\mathbf{J}_{a}$ and $\Gamma_{a5}=\Gamma_a$ are the Dirac matrices in five dimensions with the signature $\left( -,+,+,+,+\right) $. We also have the matrices $\Gamma_{ab}=\frac{1}{2}\,\left[ \Gamma _{a},\Gamma _{b}\right] $ and the ${\mathcal N}\times {\mathcal N}$ matrices $\tau _{\Lambda }$ that are generators of $su({\mathcal N})$. When ${\mathcal N}=4$, the $U(1)$ generator $\mathbf{T}_{1}$ becomes a central charge in the algebra $psu(2,2|4)$.

The supersymmetry generators $\mathbf{Q}_s^\alpha$ and $\mathbf{\bar{Q}}_\alpha^s$ carry Abelian charges $q=\pm \left(\frac{1}{4} - \frac{1}{{\mathcal N}}\right)$ and their anticommutators read
\begin{equation}
\left\{ \mathbf{Q}_{s}^{\alpha }\mathbf{,\bar{Q}}_{\beta }^{r}\right\} =
\frac{1}{4}\,\delta _{s}^{r}\,\left( \Gamma ^{AB}\right) _{\beta }^{\alpha
}\,\mathbf{J}_{AB}-\delta _{\beta }^{\alpha }\,\left( \tau ^{\Lambda
}\right) _{s}^{r}\,\mathbf{T}_{\Lambda }+i\,\delta _{\beta }^{\alpha
}\,\delta _{s}^{r}\,\mathbf{T}_{1}\,.
\end{equation}

The corresponding field-strength can be written as
\begin{equation}
\mathbf{F}=\frac{1}{\ell }\,F^{a}\mathbf{J}_{a}+\frac{1}{2}\,F^{ab}
\mathbf{J}_{ab}+F^{\Lambda }\mathbf{T}_{\Lambda }
+\left( \nabla \bar{\psi}^{s}\mathbf{Q}_{s}
-\mathbf{\bar{Q}}^{s}\nabla \psi _{s}\right) +F\,\mathbf{T}_{1}\;,
\end{equation}
where the components have the form
\begin{equation}
\begin{tabular}[b]{llll}
$F^{a}$ & $=\frac{1}{\ell }\,T^{a}+\frac{1}{2}\,\bar{\psi}^{s}\Gamma
^{a}\psi _{s}\,,\medskip $ & $F^{\Lambda }$ & $=\mathcal{F}^{\Lambda }
+\bar{\psi}^{s}\left( \tau ^{\Lambda }\right) _{s}^{r}\psi _{r}\,,\smallskip $ \\
$F^{ab}$ & $=R^{ab}+\frac{1}{\ell ^{2}}\,e^{a}e^{b}-\frac{1}{2}
\bar{\psi}^{s}\Gamma ^{ab}\psi _{s}\,,\qquad $ & $F$ & $=dA-i\bar{\psi}^{s}\psi _{s}\,.
$
\end{tabular}
\end{equation}
Here, $T^{a}$ and $R^{ab}$ are the spacetime torsion and curvature 2-forms, respectively, $\mathcal{F}^{\Lambda }$ is the field-strength 2-form for $su({\mathcal N})$, and the covariant derivative acts on fermions as
\begin{equation}
\nabla \psi _{s}=\left( d+\frac{1}{4}\,\omega ^{ab}\Gamma _{ab}
+\frac{1}{2\ell }\,e^{a}\Gamma _{a}\right) \psi _{s}-\mathcal{A}^{\Lambda }
\left( \tau_{\Lambda }\right) _{s}^{\ r}\,\psi _{r}+i\left( \frac{1}{4}
-\frac{1}{{\mathcal N}}\right) A\,\psi _{s}\,.
\end{equation}

The invariant tensor of rank three of the supergroup, completely symmetric in bosonic and antisymmetric in fermionic indices, is defined by
\begin{equation}
g_{MNK}\equiv i\left\langle \mathbf{T}_{M}\mathbf{T}_{N}
\mathbf{T}_{K}\right\rangle _{g}=\frac{1}{2}\,\text{Str\thinspace }
\left[ \left(\mathbf{T}_{M}\mathbf{T}_{N}+\left( -\right) ^{\epsilon _{M}\epsilon _{N}}
\mathbf{T}_{N}\mathbf{T}_{M}\right) \mathbf{T}_{K}\right] \,.
\end{equation}
For the particular super unitary group, its nonvanishing components are
\begin{equation}
\begin{array}{llll}
g_{[AB][CD][EF]} & =\frac{k}{2}\,\epsilon_{ABCDEF}\,,\medskip
& g_{1[AB][CD]}  & =\frac{k}{4}\,\eta _{[AB][CD]}\,, \\
g_{\Lambda _{1}\Lambda _{2}\Lambda _{3}} & =ik\,\left(\tau_{\Lambda _{1}}\tau _{\Lambda _{2}}\tau _{\Lambda _{3}}\right) \,,\medskip
& g_{1\Lambda _{1}\Lambda _{2}} & =\frac{k}{{\cal N}}\,g_{\Lambda _{1}\Lambda _{2}}\,, \\
g_{[AB] \left( _{r}^{\alpha }\right) \left( _{\beta }^{s}\right) }
& =\frac{ik}{4}\,\left( \Gamma _{AB}\right) _{\beta }^{\alpha }\delta_{r}^{s}\,,\medskip
& g_{1\left( _{r}^{\alpha }\right) \left( _{\beta}^{s}\right) } & =-\frac{k}{2}\,\left( \frac{1}{4}+\frac{1}{{\mathcal N}}\right) \delta
_{\beta }^{\alpha }\delta _{r}^{s}\,, \\
g_{\Lambda \left( _{r}^{\alpha }\right) \left( _{\beta }^{s}\right) } & =
\frac{ik}{2}\,\delta _{\beta }^{\alpha }\left( \tau _{\Lambda }\right)_{r}^{s}\,, &
g_{111} & =k\left(\frac{1}{4^2}-\frac{1}{{\mathcal N}^{2}}\right)\,,
\end{array}
\end{equation}
where the Killing metric of AdS group is $\eta _{[AB][CD]}=\eta_{AC}\,\eta _{BD}-\eta _{AD}\,\eta _{BC}$, with $\eta_{AB}=$ diag $(\eta _{ab},-)$. Similarly, $g_{\Lambda _{1}\Lambda _{2}}$ is the Killing metric of $SU({\mathcal N})$.

Having the gauge group and its symmetric invariant tensor, the Chern-Simons Lagrangian $L_{\text{CS}}(\mathbf{A})$ is implicitly defined as a five-form whose exterior derivative gives a Chern class,
\begin{equation}
dL_{\text{CS}}(\mathbf{A})=\frac{i}{3}\left\langle \mathbf{F}^{3}\right\rangle_{g}=\frac{1}{3}\,g_{MNK}\,F^{M}F^{N}F^{K}\,,  \label{dL 5}
\end{equation}
where $k$ is a dimensionless constant and the wedge symbol between forms is omitted for simplicity. The explicit expression for the CS action reads
\begin{equation}
I_{\text{CS}}\left[ \mathbf{A}\right] =\int\limits_{M}L_{\text{CS}}(\mathbf{A})=\frac{i}{3}\int\limits_{M}\left\langle \mathbf{AF}^{2}-\frac{1}{2}\,\mathbf{A}^{3}\mathbf{F}+\frac{1}{10}\,\mathbf{A}^{5}\right\rangle _{g}\,,  \label{I5}
\end{equation}
and it can be written, up to boundary terms, as
\begin{equation}
L=L_{\text{AdS}}+L_{SU({\mathcal N})}+L_{U(1)}+L_{\text{fermions}}\,.
\end{equation}
The gravitational sector of the theory is given by the Einstein-Hilbert Lagrangian with negative cosmological constant and the Gauss-Bonnet term with fixed coupling,
\begin{equation}
L_{\text{AdS}}=\frac{k}{4\ell }\,\epsilon _{abcde}\,\left( R^{ab}R^{cd}+\frac{2}{3\ell ^{2}}\,R^{ab}e^{c}e^{d}+\frac{1}{5\ell ^{4}}
\,e^{a}e^{b}e^{c}e^{d}\right) e^{e}\,.
\end{equation}
The matter sector is described by
\begin{equation}
\begin{array}{lll}
L_{SU({\mathcal N})} & = & \frac{ik}{3}\,\mbox{Tr}\left( \mathcal{AF}^{2}-\frac{1}{2}\,
\mathcal{A}^{3}\mathcal{F}+\frac{1}{10}\,\mathcal{A}^{5}\right) \,,\medskip  \\
L_{U(1)} & = & -\frac{k}{3}\,\left( \frac{1}{4^{2}}-\frac{1}{{\cal N}^{2}}\right) A(dA)^{2}+\frac{k}{4\ell ^{2}}\,\left( T^{a}T_{a}-\frac{\ell ^{2}}{2}\,R^{ab}R_{ab}-R^{ab}e_{a}e_{b}\right) A-\frac{k}{{\cal N}}\,\mathcal{F}^{\Lambda }\mathcal{F}_{\Lambda }A\,,\medskip  \\
L_{\text{fermions}} & = & -\frac{ik}{4}\,\bar{\psi}^{s}\left[ \frac{1}{\ell}\,T^{a}\Gamma _{a}+\frac{1}{2}\,\left( R^{ab}+\frac{1}{\ell ^{2}}
\,e^{a}e^{b}\right) \Gamma _{ab}+2i\left( \frac{1}{{\cal N}}+\frac{1}{4}\right)
\,dA-\bar{\psi}^{r}\psi _{r}\right] \nabla \psi _{s}\medskip  \\
&  & -\frac{ik}{2}\bar{\psi}^{s}\left( \mathcal{F}_{s}^{r}-\frac{1}{2}\,%
\bar{\psi}^{r}\psi _{s}\right) \nabla \psi _{r}+\text{c.c.}\,,%
\end{array}
\end{equation}
where $\mathcal{F}_r ^s =\mathcal{F}^\Lambda (\tau_\Lambda )_r ^s$. Supersymmetry algebra of this action closes off-shell by
construction, without addition of auxiliary fields \cite{Troncoso-Zanelli(off-shell)}.

The case ${\mathcal N}=4$ is special, because the gravitini are electrically neutral in this case and the Abelian generator becomes a central extension in the superalgebra $su(2,2|4)$, since the component $g_{111}$ vanishes. This significantly changes the dynamics of Abelian field and may produce a change in number of degrees of freedom in some backgrounds \cite{canonical-CS-sectors}.

%%%%%%%%%%%%%%%%%%%%%%%%%%%%%%%%%%%
\section{Riemann-Cartan geometry \label{Conventions}}    %   B   %
%%%%%%%%%%%%%%%%%%%%%%%%%%%%%%%%%%%
In Riemann-Cartan geometry, the vielbein $e^{a}$ and $\omega ^{ab}$ are independent fields. The spin connection, however, can be decomposed to the torsion-free connection, $\tilde{\omega}^{ab}$, that fulfills $D(\tilde{\omega})e^{a}=0$, and the contorsion, $K^{ab}=-K^{ba}$,
\begin{equation}
\omega ^{ab}=\tilde{\omega}^{ab}+ K^{ab}.
\end{equation}
The contorsion one-form $K^{ab}=K^{ab}{}_\mu \,dx^\mu$ is related to the torsion two-form $T^a =\frac{1}{2}\,T^a_{\mu\nu} \, dx^\mu \wedge dx^\nu$, by $T^a=K^a{}_b \wedge e^b$. They are in turn related to the torsion and contorsion \textit{tensors}, whose components in the coordinate basis are defined by $T^a_{\mu\nu} = e^a_\lambda T^\lambda_{\mu\nu}$, and ${K^{ab}}_\mu  =e^a_\lambda e^b_\rho {K^{\lambda \rho}}_\mu$. The following identities can be verified,
\begin{equation}
T_{\lambda \mu \nu }=K_{\lambda \nu \mu }-K_{\lambda \mu \nu }\,,\quad \text{or} \quad K_{\lambda \mu \nu } = \frac{1}{2}\,\left(T_{\mu \lambda \nu}-T_{\lambda \mu \nu }+T_{\nu \lambda \mu }\right) \,. \label{TK}
\end{equation}

If the torsion tensor is axial (i.e., totally antisymmetric), then $K_{\lambda \mu \nu }=-\frac{1}{2}\,T_{\lambda \mu \nu }$. The curvature 2-form $R^{ab}=d\omega ^{ab}+\omega _{\ c}^{a}\wedge \omega^{cb}$ can also be decomposed into the torsion-free part,$\tilde{R}^{ab} = d\tilde{\omega}^{ab}+\tilde{\omega}_{\ c}^{a}\wedge \tilde{\omega}^{cb}$, and the contorsion-dependent terms,
\begin{equation}
R^{ab}=\tilde{R}^{ab}+\tilde{D}K^{ab}+K_{\ c}^{a}\wedge K^{cb}\,.
\end{equation}
As a consequence, with the help of the identities
\begin{align}
\tilde{D}K^{ab}\wedge \tilde{D}K_{ab}& =d\left( K^{ab}\wedge \tilde{D}K_{ab}\right) -2\tilde{R}^{ab}\wedge K_a^{\ c}\wedge K_{cb}\,,  \notag \\
\tilde{R}^{ab}\wedge \tilde{D}K_{ab}& =d\left(\tilde{R}^{ab}\wedge K_{ab}\right) \,,  \notag \\
\tilde{D}K^{ab}\wedge K_a^{\ c}\wedge K_{cb}& =\frac{1}{3}\,d\left(K^{ab}\wedge K_a^{\ c}\wedge K_{cb}\right) \,,  \notag \\
K^a _{\ c}\wedge K^{cb}\wedge K_a ^{\ d}\wedge K_{db}& =0\,,
\end{align}
the Pontryagin density can be written as
\begin{equation}
R^{ab}\wedge R_{ab}=\tilde{R}^{ab}\wedge \tilde{R}_{ab}+ d\left( K^{ab}\wedge \tilde{D}K_{ab}+ 2K^{ab}\wedge \tilde{R}_{ab}+ \frac{2}{3}\,K^{ab}\wedge K_a^{\ c}\wedge K_{cb}\right) \,.  \label{RR}
\end{equation}

%%%%%%%%%%%%%%%%%%%%%%%%%%%%%%%%%%%%%%%%%%%
\section{Symmetric ansatz in AdS space \label{BH ansatz}}       %   C   %
%%%%%%%%%%%%%%%%%%%%%%%%%%%%%%%%%%%%%%%%%%%
Consider a static topological black hole ansatz in the local coordinates $x^{\mu }=(t,r,x^{m})$, $m=2,3,4$, by writing the vielbien as
\begin{equation}
e^0 =h(r)f(r)\,dt\,,\qquad e^1 =\frac{dr}{f(r)}\,,\qquad e^i =r{\hat{e}^i}= r\,\hat{e}_m^i (x)\,dx^{m}\,.  \label{e_ansatz}
\end{equation}
Here, $f$ and $h$ are arbitrary functions of the radial coordinate and ${\hat{e}^{i}}$ is the 3D vielbein of the transverse section. Without loss of generality, $f$ and $h$ can be chosen non-negative. The corresponding Levi-Civita connection
\begin{equation}
\tilde{\omega}^{01}=f(fh)' \,dt\,,\qquad \tilde{\omega}^{1i}=-f\,{\hat{e}^{i}}\,\,,\qquad \tilde{\omega}^{mn}=\hat{\omega}^{mn}\,.\label{Levi w_ansatz}
\end{equation}
In terms of the metric, this ansatz takes a familiar form,
\begin{equation}
ds^2 = g_{\mu \nu}\,dx^\mu dx^\nu = -h^2 (r) f^2 (r)\,dt^2 + \frac{dr^2}{f^2 (r)} + r^2 \gamma _{mn}(x)\,dx^m dx^n\,,  \label{BH}
\end{equation}
where the transverse metric, $\gamma_{mn}=\hat{e}_m^i \hat{e}_n^j\,\delta_{ij}$, describes a maximally symmetric 3D manifold
of unit radius, $\mathcal{R}^{mn}_{kl}(\gamma) = \kappa \delta^{[mn]}_{[kl]}$, whose geometry can be flat ($\kappa=0$), spherical ($\kappa=1$) or hyperbolic ($\kappa=-1)$. Hereafter, let us consider $\gamma_{mn}=\delta_{mn}$ ({\it i.e.} $\kappa=0$) for simplicity. Then $\hat{e}_m^i=\delta_m^i$.

The isometries of spacetime are obtained from the Killing equation
\begin{equation}
\pounds_\xi g_{\mu \nu} =\partial_\mu \xi^\lambda g_{\lambda \nu} + \partial_\nu \xi ^\lambda g_{\mu \lambda} + \xi^\lambda \partial_\lambda g_{\mu \nu} = 0\,.
\end{equation}
The general solution for a Killing vector is,
\begin{equation}
\xi=\xi ^\mu \partial_\mu  =c\,\partial_t + \frac{1}{2}\,a^{mn} (x_n \partial_m - x_m \partial_n) + b_m \partial_m\,,  \label{KV}
\end{equation}
describes the time translations, $\partial _{t}$, translations in flat directions, $\partial_m$, and spatial rotations in transverse section, $x_n \partial_m - x_m \partial_n$.
The Abelian gauge field $F=dA$ has the same isometries (\ref{KV}) if it satisfies
\begin{equation}
\pounds_\xi F_\mu \nu = \partial_\mu \xi ^\alpha F_{\alpha \nu} + \partial_\nu \xi ^\alpha  F_{\mu \alpha }+ \xi^\alpha \partial_\alpha F_{\mu \nu}=0\,,
\end{equation}
that means that its form has to be $F=F_{tr}(r)\,dt\wedge dr$. Choosing the Abelian gauge field as
\begin{equation}
A=A_t (r)\,dt\,,  \label{gauge}
\end{equation}
the field-strength reads
\begin{equation}
F=dA=-A'_{t}(r)\,dt\wedge dr\,.  \label{F}
\end{equation}

Similarly, if we require that the torsion tensor has the same isometries as a topological AdS black hole, it must satisfy
\begin{equation}
\pounds_\xi T_{\mu \nu \lambda } = \partial_\mu \xi^\alpha T_{\alpha \nu \lambda} + \partial_\nu \xi^\alpha T_{\mu \alpha \lambda} + \partial_\lambda \xi^\alpha T_{\mu \nu \alpha} + \xi^\alpha \partial_\alpha T_{\mu \nu \lambda} =0\,.
\end{equation}
Invariance under $\partial _{t}$ and $\partial _{m}$ implies that $T_{\mu \nu\lambda }$ can be a function of the radial coordinate only. Furthermore, solving the above equation gives the most general spherically symmetric torsion tensor,
\begin{equation}
\begin{array}{lll}
T_{ttr}=\chi _{t}(r)\,,\qquad  & T_{ntm}=\psi _{t}(r)\,\delta _{nm}\,,\qquad
& T_{nmk}=\phi (r)\,\epsilon _{nmk}\,. \\
T_{rtr}=\chi _{r}(r)\,, & T_{nrm}=\psi _{r}(r)\,\delta _{nm}\,. &
\end{array}
\label{T_ansatz}
\end{equation}
The torsion 2-form is then
\begin{align}
T^{0}& =-\frac{\chi _{t}}{hf}\,dt\wedge dr\,,  \notag \\
T^{1}& =f\,\chi _{r}\,dt\wedge dr\,,  \notag \\
T^{i}& =\frac{1}{r}\,\left( \psi _{t}\,dt+\psi _{r}\,dr\right) \wedge dx^{i}+
\frac{\phi }{2r}\,\,\delta ^{ik}\epsilon _{knm}\,dx^{n}\wedge dx^{m}\,.
\label{T}
\end{align}

Using the formula (\ref{TK}), we find the non-vanishing components of the contorsion,
\begin{equation}
\begin{array}{lll}
K_{trt}=\chi _{t}\,,\qquad  & K_{tnm}=\psi _{t}\,\delta _{nm}\,,\qquad  &
K_{nmk}=-\frac{1}{2}\,\phi \,\epsilon _{nmk}\,, \\
K_{trr}=\chi _{r}\,, & K_{rnm}=\psi _{r}\,\delta _{nm}\,, &
\end{array}
\end{equation}
and the contorsion 1-form,
\begin{equation}
\begin{array}[b]{llll}
K^{01} & =-\frac{1}{h}\left( \chi _{t}\,dt+\chi _{r}dr\right) \,,\medskip
& K^{1i} & =\frac{f\psi _{r}}{r}\,dx^{i}\,, \\
K^{0i} & =-\frac{\psi _{t}}{rhf}\,dx^{i}\,, & K^{ij} & =-\frac{\phi }{%
2r^{2}}\,\epsilon ^{ijk}\,\delta _{km}\,dx^{m}\,.%
\end{array}
\label{K}
\end{equation}
The full spin connection then reads
\begin{equation}
\begin{array}[b]{llll}
\omega ^{01}= & \omega \,dt-\chi dr\ ,\qquad  & \omega ^{1i}= &
\nu \,dx^{i}\ \medskip , \\
\omega ^{0i}= & -\psi \,dx^{i}\ , & \omega ^{ij}= & -\varphi %
\,\epsilon ^{ijk}\,dx_{k}\,,
\end{array} \label{spinconnection_ansatz}
\end{equation}
where we introduced new fields
\begin{equation}
\begin{array}[b]{llllll}
\omega = & f\left( fh\right) ^{\prime }-\frac{\chi _{t}}{h},\qquad  &
\nu = & \frac{f\left( \psi _{r}-r\right) }{r}\medskip ,\qquad  &
\chi  & =\frac{\chi _{r}}{h}, \\
\psi = & \frac{\psi _{t}}{rhf}\,, & \varphi =
& \frac{\phi }{2r^{2}}\,. &  &
\end{array} \label{new fields}
\end{equation}

The torsionless Riemann curvature in given ansatz has the form
\begin{eqnarray}
\tilde{R}^{01} &=&-\left(f\,(fh)' \right)' dt\wedge dr\,, \notag \\
\tilde{R}^{0i} &=&-f^2\left( fh\right)' \,dt\wedge dx^i\,, \notag \\
\tilde{R}^{1i} &=&-f' \,dr\wedge dx^i\,,  \notag \\
\tilde{R}^{ij} &=& -f^2 \,dx^i \wedge dx^j\,,  \label{RiemannBH}
\end{eqnarray}
and, consequently, the torsion-free Pontryagin density vanishes,
\begin{equation}
\frac{1}{2}\,\tilde{R}^{ab}\wedge \tilde{R}_{ab}=0\,.
\end{equation}
The torsional invariants are
\begin{align}
T^a \wedge T_a & =2r\varphi\,\left[hf\psi\,dt +\left(1+\frac{\nu}{f}\right)\,dr\right] \,\epsilon_{knm}\,dx^k \wedge dx^n \wedge dx^m\,,\notag \\
T^a \wedge e_a & =r^2\varphi\,\epsilon_{knm} \,dx^k \wedge dx^n \wedge dx^m\,.  \label{Tinv}
\end{align}
The contorsion invariants have the form
\begin{align}
\frac{1}{2}\,K^{ab}\wedge \tilde{D}K_{ab}& =-2f\varphi(f+\nu)\,\epsilon_{knm}\,dx^k\wedge dx^n\wedge dx^m\,, \notag \\
K^{ab}\wedge \tilde{R}_{ab}& =f^2 \varphi \,\epsilon_{knm}\,dx^k \wedge dx^n \wedge dx^m\,,  \notag \\
\frac{1}{3}\,K^{ab}\wedge \left( K^2\right)_{ab}& =\varphi \left[\frac{1}{3}\,\varphi ^2 +(f+\nu)^2-\psi ^2\right] \epsilon_{knm}\,dx^k\wedge dx^n\wedge dx^m\,.
\end{align}

The full Riemann curvature can be written as
\begin{align}
R^{01}& =-\omega' \,dt\wedge dr\,,  \notag \\
R^{0i}& =\omega \nu \,dt\wedge dx^i -\left( \chi \nu  +\psi ' \right) dr\wedge dx^i -\varphi \psi \,\epsilon^i_{\ jk}\,dx^j\wedge dx^k\,,  \notag \\
R^{1i}& =\left(\chi \psi +\nu ' \right) dr\wedge dx^i -\psi \omega \,dt\wedge dx^i +\varphi \,\nu \,\epsilon^i_{\ jk}\,dx^j \wedge dx^k\,,  \notag \\
R^{ij}& =-\varphi '\,\epsilon^{ij}_{\ k}\,dr\wedge dx^k +\left(\psi ^2 -\varphi ^2 -\nu ^2\right) dx^i\wedge dx^j\,.
\end{align}
It is also useful to write the Einstein-Hilbert and Gauss-Bonnet terms in Riemann-Cartan space,
\begin{align}
\epsilon _{abcde}\,R^{ab}\wedge e^{c}\wedge e^{d}\wedge e^{e}& =36\,d^{5}x\left[ -\frac{\omega ^{\prime }r^{3}}{3}+\frac{\omega \nu r^{2}}{f}%
+hfr^{2}\left( \chi \psi +\nu ^{\prime }\right)
+hr\left( \psi ^{2}-\varphi ^{2}-\nu ^{2}\right) \right]
\,,  \notag \\
\epsilon _{abcde}\,R^{ab}\wedge R^{cd}\wedge e^{e}& =24\,d^{5}x\left[
\left( -\omega ^{\prime }r+\frac{\omega \nu }{f}+hf\,\left( \chi\psi +\nu ^{\prime }\right) \right) \left( \psi ^{2}-\varphi ^{2}-\nu ^{2}\right) +\right.   \notag \\
& +\left. \omega r\,\left( \nu ^{2}-\psi ^{2}\right) ^{\prime
}-2hf\nu \,\varphi \varphi ^{\prime }\right] \,.
\end{align}
Besides, the expression for $B$ which appears in $L_{\text{int}}=\alpha\,dB\wedge A$ is given by
\begin{align}
B  & =\frac{1}{2}\,K^{ab}\tilde{D}K_{ab}+K^{ab}\tilde{R}_{ab}+\frac{1}%
{3}\,K^{ab}\left(  K^{2}\right)  _{ab}-\frac{1}{\ell^{2}}\,T^{a}e_{a}\ ,\notag \\
& =\varphi \left(  \frac{1}{3}\,\varphi ^{2}+\nu ^{2}%
-\psi ^{2}-\frac{r^{2}}{\ell^{2}}\right)  \epsilon_{knm}\,dx^{k}\wedge
dx^{n}\wedge dx^{m}\,.
\end{align}

%%%%%%%%%%%%%%%%%%%%%%%%%%%%%%%%%
\subparagraph{Levi-Civita conventions.}
%%%%%%%%%%%%%%%%%%%%%%%%%%%%%%%%%
It is useful to clarify how the three-dimensional flat transverse subspace is
embedded in the five-dimensional manifold from the point of view of its
constant tensors. The five-dimensional Levi-Civita tensor $\epsilon_{abcde}$
in the tangent space is normalized as $\epsilon_{01234}=1$. Then, the
spacetime volume form in five dimensions is given by
\begin{equation}
dx^{\mu}\wedge dx^{\nu}\wedge dx^{\alpha}\wedge dx^{\beta}\wedge
dx^{\gamma}=-\epsilon^{\mu\nu\alpha\beta\gamma}\,d^{5}x\,,
\end{equation}
what is consistent with the fact that $\epsilon^{01234}=\epsilon^{trxyz}=-1$.
In the coordinates used here, the volume element reads $d^{5}x=dt\wedge
dr\wedge dx\wedge dy\wedge dz$. Covariant Levi-Civita tensors are $\sqrt
{-g}\,\epsilon_{\mu\nu\alpha\beta}$ and $\frac{1}{\sqrt{-g}}\,\epsilon^{\mu
\nu\alpha\beta}$, where the Jacobian is $\sqrt{-g}=hr^{3}$.

On the other hand, assuming the planar horizon for the sake of simplicity, the
three-dimensional Levi-Civita tensor is
\begin{equation}
^{\left(  3\right)  }\epsilon_{mnl}\equiv\epsilon_{mnl}=\epsilon
_{trmnl}\,,\qquad\gamma_{mn}=\delta_{mn}\,.
\end{equation}
Using this notation, we find the relation between 3D and 5D tensors to be
\begin{equation}
\epsilon_{01ijk}=\epsilon_{mnl}\,\delta_{i}^{m}\delta_{j}^{n}\delta_{k}^{l}=\epsilon_{ijk}\,.
\end{equation}
We can also write the 5D volume element as
\begin{equation}
dt\wedge dr\wedge dx^{m}\wedge dx^{n}\wedge dx^{k}=\epsilon^{mnk}\,d^{5}x\,.
\end{equation}
Other examples that often appear in our calculations are
\begin{align}
\epsilon_{mnl}\,dx^{m}\wedge dx^{n}\wedge dx^{l} &  =6\,d^{3}x\,,\nonumber\\
\epsilon_{01ijk}\,\delta_{m}^{i}\delta_{n}^{j}\delta_{l}^{k}\,dt\wedge
dr\wedge dx^{m}\wedge dx^{n}\wedge dx^{l} &  =6\,d^{5}x\,.
\end{align}

%%%%%%%%%%%%%%%%%%%%%%%%%
\subparagraph{Residual gauge transformations.}
%%%%%%%%%%%%%%%%%%%%%%%%%
Gauge transformations AdS$_5 \times U(1)$ with the local parameter $\mathbf{\Lambda }=\frac{1}{\ell }\,\varepsilon^a \mathbf{J}_a +\frac{1}{2}\,\lambda^{ab}\mathbf{J}_{ab} +\theta \mathbf{T}_1\,$, act on the gauge field as $\delta \mathbf{A}= D(\mathbf{A})\mathbf{\Lambda}$ or, in components,
\begin{align}
\delta A& =d\theta \,,  \notag \\
\delta e^a & =D(\omega)\varepsilon ^a -\lambda^{ab}e_b\,,  \notag \\
\delta \omega^{ab}& =D(\omega)\lambda^{ab}+ \frac{1}{\ell^2}\,\left(e^a\varepsilon^b -\varepsilon^a e^b\right) \,.  \label{gauge tr}
\end{align}

We look for a restricted form of $\mathbf{\Lambda }$ that does not change the original spherically symmetric ansatz of the quantities defined on the spacetime manifold given by Eqs. (\ref{BH}), (\ref{T_ansatz}) and (\ref{gauge}). In other words, we want to check whether there are gauge transformations that map one spherically symmetric set of fields $\mathbf{A}(h,f,\psi ,\chi ,\phi ,A)$ into another one, $\mathbf{A}'(h', f', \psi', \chi', \phi', A')$, at the same point of spacetime.

The transformation law of the Abelian field $A$ gives
\begin{equation}
\delta A_{t}\,dt=\partial _{t}\theta \,dt+\partial _{r}\theta \,dr+\partial_{m}\theta \,dx^{m}\,,
\end{equation}
and the only transformations that fulfill this are the global ones, $\theta = const$. Thus, there is no residual $U(1)$ symmetry.

In what follows, we solve the parameters $\varepsilon^a $, $\lambda^{ab}$ assuming
\begin{equation}
h=1\,, \qquad f\neq 0 \qquad \psi \neq 0 \qquad \omega \neq 0\,.
\end{equation}
The expression for $\delta e^0$ and $\delta e^1$ written in components lead to
\begin{align}
\delta f & =\omega \,\varepsilon^1 \,,\qquad \lambda^{01}=\frac{\omega}{f} \,\varepsilon^0\,,  \notag \\
\lambda^{0i} & = -\psi \,\varepsilon^i \,,\qquad \lambda^{1i}=\frac{\eta}{r^2}\,\varepsilon^{i}\,,
\end{align}
and the conditions on local parameters,
\begin{align}
0& =\omega \,\varepsilon^1 + f^2 (\partial_r \varepsilon^1 -\chi_r \,\varepsilon^0 ) \,,  \notag \\
0& =\omega \,\varepsilon^0 - f^2 (\partial_r \varepsilon^0 - \chi_r \,\varepsilon^1) \,.  \label{system}
\end{align}
Writing also $\delta e^i$ in components gives that $\varepsilon^0$ and $\varepsilon^1$ are not independent,
\begin{equation}
\lambda^{0i}\,,\varepsilon^i,\lambda^{1i},\lambda^{im}=0\,,\qquad \varepsilon^0=-\frac{f\eta}{\psi_t}\,\varepsilon^1\,. \label{epsilon_0}
\end{equation}
Writing out the transformation law for $\delta \omega^{ab}$ in components and using previous solutions, we find
\begin{align}
\delta \omega & =\frac{f}{\ell^2}\,\varepsilon^1\,,  \notag \\
\delta \chi_r& =- \partial_r \left[f \left(\partial_r \varepsilon^0 - \chi_r \,\varepsilon^1 \right) \right] + \frac{\varepsilon^0}{\ell^2 f}\,.
\end{align}
Finally, the transformation laws $\delta \omega^{0i}$, $\delta \omega^{1i}$ and $\delta \omega^{ij}$ give
\begin{align}
\delta \psi_t& =f^2\eta \partial_r\varepsilon^0 +\left(\frac{\psi_t}{f}\omega \,-f^2 \eta \chi_r \right) \,\varepsilon^1 +\frac{r^2}{\ell^2}\,f\varepsilon^0\,,  \notag \\
\delta \eta & =\psi_t\left(\partial_r\varepsilon^0-\chi_r\,\varepsilon^1\right) - \frac{r^2}{\ell^2}\,\varepsilon^1\,, \notag \\
\delta \phi & =0\,.
\end{align}

In sum, we have to solve the system (\ref{system}) for $\varepsilon^1$, and if non-vanishing, it induces two other non-vanishing parameters: $\varepsilon^0$ and $\lambda ^{01}$, both functions of $\varepsilon^1$.

Let us solve the system (\ref{system}). One solution that always exists is $\varepsilon^1=0$, meaning that there are no residual gauge symmetries. The other possibility is $\varepsilon^1\neq 0$, but this is not possible generically, but only for particular solutions of the fields, and then $\varepsilon^1$ is completely (globally) determined, that means that there are no remaining local symmetries in this theory.

In any case, there are no local transformations that leave the ansatz invariant.

%%%%%%%%%%%%%%%%%%%%%%%%%%%%%%%%%%%%%%%%%%%%%%%%%
\section{Constraint structure of the effective action with $h\neq 1$  \label{with h}}    %     D    %
%%%%%%%%%%%%%%%%%%%%%%%%%%%%%%%%%%%%%%%%%%%%%%%%%
We consider the metric components $g_{tt}=-(hf)^2$ and $g_{rr}=1/f^2$ as independent dynamical fields. The generalized coordinates are $q_s=\left\{ f,h,A_t, \varphi ,\psi , \nu ,\omega ,\chi \right\}$ and the effective spherically symmetric CS action is given by
\begin{align}
\mathcal{I}_{\text{eff}}[q,\dot q]& =\frac{6k}{\ell }\int dr\,\left[ \left( \frac{\omega \nu }{f}+hf\chi \psi +hf\nu ^{\prime }+
\frac{hr}{\ell ^{2}}+\omega \right) \left( \psi ^{2}-\varphi ^{2}-
\nu ^{2}+\frac{r^{2}}{\ell ^{2}}\right) \right.   \notag \\
& -\left. 2\left( r\omega +hf\nu \right) \,\varphi \varphi'-\frac{\alpha\ell }{k}\,\varphi \left( \frac{1}{3}\,\varphi^{2}+\nu ^{2}-\psi ^{2}-\frac{r^{2}}{\ell ^{2}}\right)
A_{t}^{\prime }\right] \,.  \label{Ieff(h)}
\end{align}
We proceed similarly as in Section \ref{Hamiltonian analysis}. The primary constraints obtained from $\mathcal{I}_\text{eff}$ read
\begin{equation}
\begin{array}{llll}
C_{f} & =p_{f}\approx 0\,, & C_{\omega } & =p_{\omega }\approx 0\,,\medskip
\\
C_{h} & =p_{h}\approx 0\,, & C_{\nu } & =p_{\nu }-\frac{6k}{\ell }\,hf\,
\mathcal{T}_{1}\approx 0\,,\medskip  \\
C_{\psi } & =p_{\psi }\approx 0\,, & C_{\varphi } & =p_{\varphi }+\frac{12k}{\ell}\,\left( r\omega +hf\nu \right) \,\varphi \approx 0\,,\medskip
\\
C_{\chi } & =p_{\chi }\approx 0\,,\qquad  & C_{A} & =p_{A}-4\alpha\,\varphi ^{3}-6\alpha\,\varphi \mathcal{T}_{1}\approx 0\,,
\end{array}
\label{primary}
\end{equation}
and the canonical and total Hamiltonian have the form
\begin{align}
\mathcal{H}_C & \approx -\frac{6k}{\ell }\,\left( \frac{\omega \nu }{f}%
+hf\chi \psi +\frac{hr}{\ell ^{2}}+\omega \right) \mathcal{T}
_{1}\,, \notag \\
\mathcal{H}_T & = \mathcal{H}_C + u^sC_s\,,
&  \label{canonical}
\end{align}
respectively. For nonvanishing torsion, $fh\phi \psi  \chi  \neq 0$, the consistency conditions for the primary constraints give rise to only one secondary constraint,
\begin{equation}
\mathcal{T}_{1}\approx 0\,,
\end{equation}
and four out of eight multipliers become determined,
\begin{align}
u^{\varphi }& =0\,,  \notag \\
u^{A}& =-\frac{k}{\alpha\ell }\,\varphi ^{-1}\left( \frac{\omega \nu }{f}
+hf\chi \psi +\frac{hr}{\ell ^{2}}+\omega +hf\,u^{\nu}\right) \,,  \notag \\
u^{\psi }& =\psi ^{-1}\left( \nu u^{\nu }-\frac{r}{\ell ^{2}}\right) \,,  \notag \\
u^{\omega }& =\frac{1}{r}\left( \frac{\omega \nu }{f}+hf\chi
\psi +\frac{hr}{\ell ^{2}}\right) -\frac{1}{r}\left( f\nu u^{h}+h\nu u^{f}\right) \,.
\end{align}
The separation between first and second class constraints is achieved through redefinition of the constraints $C_s\rightarrow(G_a,S_\alpha)$,
\begin{align*}
\text{First class}: & \qquad  G_{a}=\left\{
G_{h},G_{f},G_{\nu },G_{\chi },G_{\tau }\right\}, \\
\text{Second class}: & \qquad S_{\alpha }=\left\{ S_{\varphi },S_{\omega },S_{\psi },S_{A}\right\}\,,
\end{align*}
where the constraints are redefined in a way that do not change the constraint surface,
\begin{equation}
\begin{array}{llll}
G_{h} & =h\left( C_{h}-\frac{f\nu }{r}\,C_{\omega }\right) \,, &
S_{\varphi } & =\varphi C_{\varphi }\,, \\
G_{f} & =f\left( C_{f}-\frac{h\nu }{r}\,C_{\omega }\right) \,, &
S_{\omega } & =\frac{hf}{r}\,C_{\omega }\,, \\
G_{\tau } & =hf\,\left(- \frac{6k}{\ell }\,\mathcal{T}_{1}-\frac{k}{\alpha\ell }
\,\varphi ^{-1}C_{A}+\frac{1}{r}\,C_{\omega }\right) \,,\quad  &
S_{\psi } & =\frac{1}{\psi }\,C_{\psi }\,, \\
G_{\nu } & =C_\nu +\frac{\nu }{\psi }\,C_{\psi}-\frac{k}{\alpha\ell}\,\frac{hf}{\varphi }\,C_{A} \,, & S_{A} & =C_{A}\,. \\
G_{\chi } & =C_{\chi }\,, &  &
\end{array}
\end{equation}
It can be checked that the Jacobian of these transformations is nonvanishing for
$fh\phi \psi  \chi  \neq 0$. The first class constraints close the algebra
\begin{align}
\left[ G_{h},G_{\nu }\right] & =-G_{\tau }\,,\qquad \left[ G_{\tau },G_{h}
\right] =G_{\tau }\,,  \notag \\
\left[ G_{f},G_{\nu }\right] & =-G_{\tau }\,,\qquad \left[ G_{\tau },G_{f}
\right] =G_{\tau }\,,
\end{align}
and the second class constraints have invertible symplectic form $\Omega_{\alpha \beta }=\{S_{\alpha },S_{\beta }\}$,
\begin{align}
\left[ S_{\varphi },\ S_{\omega }\right] & =\frac{12k}{\ell }\,hf\,
\varphi ^{2}\,,  \notag \\
\left[ S_{\varphi },\ S_{A}\right] & =6\alpha\varphi \,\mathcal{T}_{1}\,,
\notag \\
\left[ S_{\psi },\ S_{A}\right] & =12\alpha\varphi \,.
\end{align}
The rest of commutators vanish on the constraint surface,
\begin{align}
\left[ S_{\omega },\ G_{f}\right] & =S_{\omega }\,,  \notag \\
\left[ S_{\omega },\ G_{h}\right] & =S_{\omega }\,,  \notag \\
\left[ S_{\varphi },G_{\nu }\right] & =G_{\tau }-S_{\omega }\,, \\
\left[ S_{\varphi },G_{\tau }\right] & =G_{\tau }-S_{\omega }\,.  \notag
\end{align}

The symplectic matrix of these constraints can be transformed (e.g., by taking differences $G_\nu -G_\tau $ and $G_h-G_f$) into the one equivalent to setting $h=1$, because the generator that appears due to dynamical field $h(r)$, that is $G_h-G_f$, commutes with all other generators and, therefore, contributes as zero column (row) in the symplectic matrix. This generator corresponds to an Abelian symmetry that can can be gauge fixed by $h=1$. This gauge fixed system identically matches the one obtained by setting $h=1$ directly in the action. This is why, for the sake of simplicity, we start from $h=1$ in Subsection \ref{Constraints}.

%%%%%%%%%%%%%%%%%%%%%%%%%%%%%%%%%%%%%%%%%%%%%%%%%
\section{Black hole mass} \label{Mass RT}
%%%%%%%%%%%%%%%%%%%%%%%%%%%%%%%%%%%%%%%%%%%%%%%%%

So far, we have neglected all boundary terms in the Hamiltonian first order
action $I_\text{eff}= \text{Vol}(\partial\mathcal{M})\,\mathcal{I}_\text{eff}$. The boundary $\partial\mathcal{M}$ is a time-like
surface of the form $\mathbb{R}\times\gamma_3$, where $\gamma_3$ is the flat transversal section.

In order to connect the integration
constants $\mu$, $b$ and $C$ with the conserved charges, i.e., the mass of
the black hole $M$, we have to supplement the action (\ref{Ieff}) by a boundary
term $B$ defined at $r\rightarrow \infty$,
\begin{equation}
\mathcal{I}_B=\mathcal{I}_\text{eff}+B\,. \label{IB}
\end{equation}
In the Hamiltonian approach, the black hole mass is related to the boundary
terms chosen so that $\mathcal{I}_B$ has an extremum on-shell \cite{Regge-Teitelboim}.
In practice, it means that $\delta \mathcal{I}_B$ has
to vanish on-shell when the fields are kept fixed on the boundary.
Varying Eqs.(\ref{IB}) and (\ref{Ieff}) with $\omega =f(fh)'-\chi_t$ and $k=-\ell^3/16\pi G$ leads to
\begin{align}
\delta \mathcal{I}_B & =-\frac{3\ell^2}{8\pi G }\int dr\,
\frac{d}{dr}\left[ \rule{0pt}{15pt}\left( (\nu+f) \,\delta (fh)+hf\,\delta
\nu\right) \mathcal{T}_1-2rf\,\varphi \varphi'\,\delta (fh)\right.   \nonumber \\
- & \left. ( r\omega +hf\nu) \,\delta \varphi^2 -\frac{\alpha\ell }{k}\,\varphi \left(\frac{4}{3}\,\varphi^2
-\mathcal{T}_1\right) \delta A_t \right] + \delta B +\text{e.o.m.}\,,
\end{align}
where e.o.m. are the bulk terms that vanish using the equations of motion. In order to have
$\delta \mathcal{I}_B=0$ on-shell, we find that the boundary term has to satisfy
\begin{align}
\delta B & =\frac{3\ell^2}{8\pi G }\,\lim_{r\rightarrow \infty }\left[ \rule{0pt}{15pt}
\left( (\nu+f) \,\delta (fh)+hf\,\delta \nu\right) \mathcal{T}_1\right.   \nonumber \\
& -\left. 2rf\,\varphi \varphi'\,\delta (fh)-(r\omega +hf\nu) \,\delta \varphi^2
-\frac{\alpha \ell }{k}\,\left( \frac{4}{3}\,\varphi^3-\varphi\mathcal{T}_1\right)
\delta A_t\right] \,.  \label{var_B}
\end{align}

The mass $M(\mu ,b,C)$ can be obtained only if one is able to integrate out the variation $\delta M(\mu ,b,C)=\text{Vol}(\gamma_3)\,\delta B(\mu ,b,C)$ for a particular solution.
To analyze it better, let us look at the black hole solutions discussed in Sections \ref{Charged BH ansatz} and \ref{Comparison}.

\begin{quote}
(\textit{i}) \textit{Uncharged, static black hole without torsion}
\end{quote}

The mass for this CS black hole was calculated in Ref.\cite{Aros-Zanelli}.
The non-vanishing fields are $h=1$ and $f=-\nu=\sqrt{\frac{r^2}{\ell^2}-\mu }$,
and the function $\mathcal{T}_1=-f^2+\frac{r^2}{\ell^2}$ is non-vanishing. The mass is obtained from
\begin{equation}
\delta M(\mu )=\frac{3\ell^2\,\text{Vol}(\gamma_3)}{16\pi G }\,\lim_{r\rightarrow \infty }\left(f^2-\frac{r^2}{\ell^2}\right) \delta f^2
=\frac{3\ell^2\,\text{Vol}(\gamma_3)}{16\pi G }\,\mu \delta \mu \,,
\end{equation}
that can be easily integrated out. We arrive to the result
\begin{equation}
M=\frac{3\ell^2\,\text{Vol}(\gamma_3)}{32\pi G }\,\mu^2\,,
\end{equation}
where the integration constant is chosen so that the mass vanishes for the vacuum solution.
This result matches the one found in Ref.\cite{Aros-Zanelli}.

\begin{quote}
(\textit{ii}) \textit{Static, charged black hole with two components of torsion}
\end{quote}

Here we calculate the mass for the black hole solution found in Subsection \ref{Charged BH sol},
which has only the axial torsion $T_{nmk}=\phi\,\epsilon_{nmk}$ and the torsion components $T_{nrm}=\psi_r\,\delta_{nm}$ non-vanishing.
They correspond to the fields $\varphi=C$ and $\nu=-\sqrt{\frac{r^2}{\ell^2}-C^2}$.
The metric functions are $h=1$ and $f=\sqrt{\frac{r^2}{\ell^2}+br-\mu}$, and the electric potential at the infinity is
$A_t(\infty)=\Phi-\frac{1}{2}\left(\mu +C^2+\frac{1}{4}\,b\ell^2\right)$.

In this case, the first term in Eq.(\ref{var_B}) that was dominant in the torsionless case now vanishes
due to $\mathcal{T}_1=0$, leading to
\begin{equation}
\delta M(\mu ,b,C)=-\frac{3\ell^2\,\text{Vol}(\gamma_3)}{8\pi G }\,\lim_{r\rightarrow \infty }\left[(rff'+f\nu) \,\delta C^2
+\frac{4\alpha\ell}{3k}\,C^3\delta A_t\right] \,,
\end{equation}
where we replaced $\ell \rightarrow -\ell $ in this branch of the solution.
Keeping the electric potential zero at the boundary, we find
\begin{equation}
\delta M(\mu ,b,C)=\frac{3\ell^2\,\text{Vol}(\gamma_3)}{16\pi G }\left(C^2+\mu +\frac{b^2\ell ^2}{4}\right) \,\delta C^2\,.
\end{equation}

The mass $M(\mu ,b,C)$ cannot be integrated out for general choice of the
integration constants. In the particular point in the space of parameters
when $C^2=\mu +\frac{b^2\ell^2}{4}$, then the electric potential
loses the $1/r$ term in the asymptotic expansion (\ref{asymptotic At}) and
the electromagnetic field energy becomes finite. Then the mass can be integrated to give
\begin{equation}
M=\frac{3\ell^2\,\text{Vol}(\gamma_3)}{16\pi G }\,\left(\mu +\frac{b^2\ell^2}{4}\right)^2\,.
\end{equation}
When $b=0$, the mass is a double of the torsionless one, that shows that these solutions are not equivalent.
\begin{quote}
(\textit{iii}) \textit{Static, uncharged black hole with axial torsion}
\end{quote}

When only the axial torsion, $\varphi=C$, is present \cite{Canfora:2007xs}, we have $h=1$,
$f=-\nu=\sqrt{\frac{r^2}{\ell^2}-\mu}$ and $\mathcal{T}_1=-C^2-f^2+\frac{r^2}{\ell^2}$.
Then
\begin{equation}
\delta M(\mu,C)=\frac{3\ell^2\,\text{Vol}(\gamma_3)}{32\pi G }\,\lim_{r\rightarrow \infty }
\left[ \rule{0pt}{15pt} \delta (\mu C^2-\mu^2)+2\mu\,\delta C^2\right] \,,
\end{equation}
and the mass cannot be integrated out in general. Again we set $\ell \rightarrow -\ell $ in this branch.
At the particular point $C^2=\mu$, the mass is
\begin{equation}
 M=\frac{3\ell^2\,\text{Vol}(\gamma_3)}{16\pi G }\,\mu^2 \,.
\end{equation}

%%%%%%%%%%%%%%%%%%%%%%%%%%%%


\begin{thebibliography}{99}
\bibitem{SO(4-2)} Hereafter, we will refer to $SO(4,2)$ group as AdS$_5$ group.
\end{thebibliography}
\end{document}